\newif\ifloguseIEEEConf
\ifloguseIEEEConf
    \documentclass[10pt,conference]{IEEEtran}
\else
    \documentclass{article}
\fi

\usepackage{graphicx}
\usepackage{lscape}
\usepackage{amssymb}
\usepackage[letterpaper,top=3cm,bottom=3.25cm,left=2.75cm, right=2.75cm]{geometry}
\usepackage{amsfonts,amsmath, amssymb, graphics,geometry,framed,enumitem}
\usepackage{arydshln}
\begin{document}

\newtheorem{theorem}{Theorem}
\newtheorem{corollary}{Corollary}
\newtheorem{lemma}{Lemma}
\newtheorem{example}{Example}
\newtheorem{definition}{Definition}
\newtheorem{proposition}{Proposition}
\newtheorem{observation}{Observation}
\newtheorem{conjecture}{Conjecture}
\newtheorem{remark}{Remark}
\newcommand{\N}{\mathbb{N}}
\newcommand{\qed}{\hfill $\diamondsuit$}
\newcommand{\noam}{Noam Presman\ }
\newcommand{\bhat}{Bhattacharyya }
\newcommand{ \etld } {\tilde{\epsilon}}
\newcommand{ \He } {\mathcal{H}}

\ifloguseIEEEConf

\else
  \newcommand{\QED}{\hfill $\diamondsuit$}
  \newcommand{\proof}{\noindent {\bf Proof}\ }
\fi

\title{Binary Polarization Kernels from Code Decompositions  }

\author{Noam Presman, Ofer Shapira, Simon Litsyn\thanks{Noam Presman, Ofer Shapira and Simon Litsyn are with the
   the School of Electrical Engineering,
   Tel Aviv University, Ramat Aviv 69978 Israel.
   (e-mails:
 \{presmann, ofershap, litsyn\}@eng.tau.ac.il.).},
Tuvi Etzion\thanks{Tuvi Etzion is with the
   Department of Computer Science,
   Technion, Haifa 32000, Israel.
   (e-mail: \texttt{etzion@cs.technion.ac.il}).},
   and Alexander Vardy\thanks{Alexander Vardy is with the
   Department of Electrical and Computer Engineering and
   the Department of Computer Science and Engineering,
   University of California San Diego,
   La Jolla, CA 92093--0407, U.S.A.
  (e-mail: \texttt{avardy@ucsd.edu}). \newline
   {The work of Tuvi Etzion and Alexander Vardy
  was supported in part by the United States --- Israel
Binational Science~Foundation (BSF), Jerusalem, Israel, under Grant 2012016.}\newline{This paper was presented in part at the 2011 IEEE International Symposium on Information Theory,  Saint Petersburg, Russia. A pre-print of some of the results is also available at http://arxiv.org/abs/1101.0764.}\newline{Copyright (c) 2014 IEEE. Personal use of this material is permitted.  However, permission to use this material for any other purposes must be obtained from the IEEE by sending a request to pubs-permissions@ieee.org.}}}
\date{}

\maketitle

\begin{abstract}
In this paper, code decompositions (a.k.a.\ code nestings) are used
to design binary polarization kernels. The proposed kernels are
in general non-linear. They provide a better polarization exponent
than the previously known kernels of the same dimensions. In particular,
non-linear kernels of dimensions $14$, $15$, and $16$ are constructed and are
shown to have optimal asymptotic error-correction performance.
The optimality is proved by showing that the exponents of these
kernels achieve a new upper bound that is developed in this paper.
\end{abstract}

\section{Introduction}
Polar codes were introduced by Arikan \cite{Arikan} and provided
a scheme for achieving the symmetric capacity of binary
memoryless channels (B-MC) with polynomial encoding and decoding
complexities. Arikan used  a simple construction based on the
following linear kernel
$$
G_2 = \left[
      \begin{array}{cc}
        1 & 0 \\
        1 & 1 \\
      \end{array}
    \right].
$$
In this scheme,  a $2^n\times2^n$ matrix, $G_2^{\otimes n}$, is generated by performing the  Kronecker power on $G_2$. An input vector $\bf u$ of length   $N=2^n$  is transformed into an $N$ length vector $\bf x$ by multiplying a certain permutation of the vector $\bf u$ by $G_2^{\otimes n}$. The vector $\bf x$ is  transmitted  through  $N$ independent copies of the memoryless channel, $\mathcal{W}$.  This results in  $N$ new (dependent) channels between the individual components of $\bf u$  and the outputs of the channels. Arikan showed that these
channels exhibit the phenomenon of polarization under successive
cancelation (SC) decoding. This means that as $n$ grows, there is a
proportion of $I(\mathcal{W})$ (the symmetric channel capacity) of the
channels that become clean channels (i.e. having the capacity
approaching $1$) and the rest of the channels become completely
noisy (i.e. with the capacity approaching $0$). An important
question is how fast the polarization occurs in terms of the code's
length $N$. Arikan and Telatar \cite{Arikan2} analyzed the rate of polarization
for the $2\times 2$ kernel, and showed that the rate is
$O\left(2^{-N^{0.5}}\right)$. More precisely,  they proved that
\begin{equation}\label{eq:introRate1}
\liminf_{n\rightarrow \infty}\Pr\left(Z_n\leq 2^{-N^\beta}\right)=I(\mathcal{W})
\,\,\,\, \text{for}\,\,\, \beta<0.5
\end{equation}
\begin{equation}\label{eq:introRate2}
\liminf_{n\rightarrow \infty}\Pr\left(Z_n\geq 2^{-N^\beta}\right)=1
\,\,\,\,\text{for}\,\,\,\, \beta>0.5, \end{equation}
where $\left\{Z_n\right\}_{n\geq 0}$ is the \bhat parameter random sequence corresponding to Arikan's random tree process \cite{Arikan}.

 Korada \textit{et al.}  \cite{Korada} studied the use of alternatives
to $G_2$ for the symmetric B-MC. They gave necessary and sufficient conditions for
polarization when linear binary kernels are used over the symmetric
B-MC channels. Furthermore,    the notion of the rate of
polarization was generalized  for linear polar codes with kernels that are defined by an arbitrary
 generating matrix $G$ of  ${\ell} \times {\ell}$ dimensions.
The rate of polarization was quantified by the  exponent of the
kernel $E(G)$, which plays the general role of the threshold (equal
$0.5$) appearing in (\ref{eq:introRate1}) and
(\ref{eq:introRate2}) (note, that here $N={\ell}^n$).  Korada \textit{et al.}
showed that $E(G)\leq 0.5$ for all binary linear kernels of
 ${\ell}\leq 15$ dimensions, which is the kernel exponent found for
Arikan's $2\times 2$ kernel, and that for ${\ell}=16$ there exists a
kernel $G$ for which $E(G)= 0.51828$, and this is the
maximum exponent achievable by a binary linear kernel up to this
number of dimensions. Furthermore, for optimal linear
kernels, the exponent $E(G)$ approaches 1 as ${\ell}\rightarrow
\infty$.

Mori and Tanaka  \cite{Mori2010} considered the general case
of a mapping $g(\cdot)$, which is not necessarily linear and binary,
as a basis for channel polarization constructions. They gave
sufficient conditions for polarization and generalized the exponent
for these cases. It was demonstrated that  non-binary,
however linear, kernels based on Reed-Solomon codes and Algebraic
Geometry codes have far better exponents
than the exponents of the known binary kernels  \cite{MoriandTanka3}. This is true even
for the Reed-Solomon kernel $G$ with ${\ell}=4$ dimensions and with alphabet
size $q=4$, in which $E\left(G\right)=0.57312$.

In this paper, we propose designing    binary kernels
having a large exponent, by using code decompositions
(a.k.a code nestings). The developed kernels show better exponents
than the ones considered by Korada \textit{et al.}  \cite{Korada} for the same number dimensions.  Moreover, we describe binary non-linear kernels of  $14$, $15$ and $16$ dimensions providing  superior exponents
 than any binary linear kernel of the same number of dimensions.

The paper is organized as follows. In Section
\ref{sec:CodesDecom}, we describe kernels  that are constructed using
decompositions of codes into sub-codes. Furthermore, by using Mori and Tanaka's results on the exponent \cite{Mori2010}, we observe that the exponent of these kernels is a function of the partial minimum distances between the sub-codes.  We then develop in Section \ref{sec:Bounds} an upper-bound on the exponent of a kernel with $\ell$ dimensions. In Section
\ref{sec:ExmpOfGoodKernels}, we give examples of known code decompositions which result in binary kernels that achieve  upper-bounds from Section \ref{sec:Bounds}. In Section \ref{sec:ExmpofModifiying}, we give lower-bounds on the exponent for kernel of  $18\leq\ell\leq 25$ dimensions. These lower-bounds are derived based on modifications of classical code constructions.

This paper is an extended version of our conference paper \cite{Presman2011a}. The main additional contributions in this version are: \textbf{(i)} A Detailed description and a proof of a new upper bound on the optimal exponent of kernels (which is also valid for non-binary cases) in Section \ref{sec:Bounds}. Using this bound we prove that the non-linear kernels introduced in \cite{Presman2011a} are indeed optimal. \textbf{(ii)} Description of seven new binary linear kernels having the largest known exponent per their dimensions in Section \ref{sec:ExmpofModifiying}.


Employing code decompositions for generating good polar codes is also used by the authors to construct \textit{mixed-kernels} codes \cite{Presman2011}. However, as opposed to this paper in which the kernels are binary and induced by code decompositions, in  \cite{Presman2011} we allow the decomposition steps to be of various sizes, thereby allowing more powerful polar code structures.

\section{Preliminaries}\label{sec:CodesDecom}
Throughout we use the following notations. For a natural number ${\ell}$, we denote
$[{\ell}]=\left\{1,2,3,...,{\ell}\right\}$ and $[{\ell}]_{-}=\left\{0,1,2,...,{\ell}-1\right\}$. We denote vectors in bold letters.  For $i\geq j$, let ${\bf
u}^i_j=\left[u_j \,\,\, u_{j+1} \ldots \,\,\,\, u_i\right]$ be the sub-vector of ${\bf u}$ of length
$i-j+1$ (if $i<j$ we say that ${\bf
u}^i_j=[\,\,\,]$, the empty vector, and its length is $0$). For two vectors $\bf u$ and $\bf v$ of lengths $n_u$ and $n_v$, we denote the $n_u+n_v$ length vector which is the concatenation of $\bf u$ to $\bf v$ by $[{\bf u} \,\,\,{\bf v}]$ or  ${\bf u}\bullet{\bf v}$  or just ${\bf u}{\bf v}$. For a scalar $x$, the  $n_u+1$  length vector ${\bf u}\bullet x$, is just the concatenation of the vector ${\bf u}$ with the length one vector containing $x$. In all of our tables, fractional numerical values are trimmed to their first five digits after the decimal point.

We consider  kernels that are based on bijective binary
transformations. A channel polarization kernel of ${\ell}$ dimensions,
denoted by $g(\cdot)$, is a bijective mapping
$$
g:\left\{0,1\right\}^{{\ell}}\rightarrow \left\{0,1\right\}^{{\ell}},
$$
i.e.  $g({\bf u})={\bf x}, \,\,\,\, {\bf u}, {\bf x}
\in\left\{0,1\right\}^{{\ell}}$. The number $\ell$ (the number of dimensions) is also referred to as the size of the kernel. Denote the output components of the
transformation by
$$
g_i({\bf u})=x_i,      \,\,\,\,\,\,\,\,\ i\in[{\ell}]_{-}.
$$
It is convenient to denote by $g^{({\bf
v}_0^{i-1})}:\left\{0,1\right\}^{{\ell}-i}\rightarrow
\left\{0,1\right\}^{{\ell}}$, the restriction of $g(\cdot)$ to the
set $$\left\{{\bf v}_0^{i-1} {\bf u}_0^{{\ell}-1-i}| {\bf
u}_0^{{\ell}-1-i}\in \left\{0,1 \right\}^{{\ell}-i}\right\},$$ that is
$$
g^{({\bf v}_0^{i-1})}({\bf u}_0^{{\ell}-1-i}) =g({\bf v}_0^{i-1} {\bf
u}_0^{{\ell}-1-i}), \,\,\,\,\,\,\,\,\,\,\, i \in [{\ell}+1]_{-}.
$$
 Next, we consider code decompositions. In this procedure, the initial code is
partitioned into several sub-codes having the same size. Each of these
sub-codes can be further partitioned. Here, we choose as the initial
code, the total space of  length ${\ell}$ binary vectors, and denote
it by $T_0^{()}=\left\{0,1\right\}^{{\ell}}$. This set is
partitioned into $m_0$ equally sized sub-codes
$T_1^{(0)},T_1^{(1)},...,T_1^{(m_0-1)}$, and  each sub-code
$T_1^{(b_0)}$ is in turn partitioned into $m_1$ equally sized codes
$T_2^{([b_0 \,\,\, 0])},T_2^{([b_0 \,\,\, 1])},...,T_2^{([b_0 \,\,\, (m_1-1)])}$
(where $b_0\in[m_0]_{-}$). This partitioning may be
further carried on.

\begin{definition} \label{def: compoIntoCosets}
The set $\left\{T_0,...,T_{m-1}\right\}$ is called a decomposition of $\left\{0,1\right\}^{\ell}$ , if
 $T_0^{()} = \left\{0,1 \right\}^{\ell}$, and $T_i^{({\bf b}_0^{i-1})}$ is partitioned into $m_i$ equally sized sets $\left\{T_{i+1}^{({\bf b}_0^{i-1}\bullet b_i)}\right\}_{b_i \in [m_i]_{-}}$, of size $\frac{2^{{\ell}}}{\prod_{j=0}^{i}m_j}$ ($i\in[m]_{-}$). We denote the set of sub-codes of level number $i$ by $T_i$, that is $$T_i = \left\{ T_i^{({\bf b}_0^{i-1})}|b_j\in [m_j]_{-} ,\,\,\,j\in [i]_{-}\right\}.$$ The partition is usually  described by the following chain of codes parameters $$(\ell,k_0,d_0)-(\ell,k_1,d_1)-...-(\ell,k_{m-1},d_{m-1}),$$ if for each $\mathcal{T}\in T_i$ we have that  $\mathcal{T}$ is a code of length $\ell$, size $2^{k_i}$ and minimum distance at least $d_i$.

 If the sub-codes of the decompositions are cosets, then we say that $\left\{T_0,...,T_{m-1}\right\}$ is a decomposition into cosets. In this case, for each $T_i$ the sub-code that contains the zero codeword is called the representative sub-code, and a minimal weight codeword for each coset  is called a coset leader. If all the sub-codes in the decomposition are cosets of linear codes, we say that the decomposition is linear.
\end{definition}
\begin{example}
Consider ${\ell}=4$ and the $4\times 4$ binary matrix
$$
G = \left[
      \begin{array}{cccc}
        1 & 0 & 0 & 0 \\
        1 & 1 & 0 & 0 \\
        1 & 0 & 1 & 0 \\
        1 & 1 & 1 & 1 \\
      \end{array}
    \right].
$$
A partition into cosets, having the following chain of parameters
$(4,4,1)-(4,3,2)-(4,1,4)$ is implied by the rows of the matrix. This is
done by taking $T_0^{()} = \left\{0,1\right\}^4$ (the code that is spanned by all the rows of the matrix), which is
partitioned into the even weight codewords and odd weight codewords
cosets  (the code generated by the three bottom rows of the matrix and its coset, that is shifted by the first row), i.e. $T_1^{(0)}=\left\{{\bf x}_0^3| \sum_{i=0}^{3}x_i \equiv
0 (\mbox{\rm mod} \, 2) \right\}$, $T_1^{(1)}=\left\{{\bf x}_0^3|
\sum_{i=0}^{3}x_i \equiv 1 (\mbox{\rm mod} \, 2) \right\}$. These
cosets are in turn partitioned into anti podalic pairs (the code generated by the last row of the matrix and its cosets),
$T_2^{([0\,\,\, 0])}=\{[0\,0\,0\,0],[1\,1\,1\,1]\}$, $T_2^{([0\,\,\,1])}=\{[1\,0\,1\,0],[0\,1\,0\,1]\}$,
$T_2^{([0\,\,\,2])}=\{[1\,1\,0\,0],[0\,0\,1\,1]\}$, $T_2^{([0\,\,\,3])}=\{[0\,1\,1\,0],[1\,0\,0\,1]\}$, and
$T_2^{([1 \,\,\, b])}=[1\, 0\, 0\, 0 ]+T_2^{([0\,\,\,b])}$ ($b\in[4]_{-}$).
Note that in order to describe this partition, it suffices to
describe the representative sub-codes and the coset leaders for the partition
of the representative sub-codes.
\end{example}
A binary transformation can be associated to a code decomposition in the following way.
\begin{definition}\label{def: transAssiciatedTheCodes}
Let $\left\{T_0,T_1,...,T_{{\ell}}\right\}$ be a code decomposition of $\left\{0,1\right\}^{\ell}$, such that $m_i=2$ for each $i\in[{\ell+1}]_{-}$. Note that the code $T_i^{\left({\bf b}_0^{i-1}\right)}$ is of size $2^{{\ell}-i}$, and specifically  $T_{{\ell}}^{{([b_0\,\,\,b_1\,\,\, \ldots\,\,\, b_{{\ell}-1}])}}$ contains only one codeword. We call such a decomposition a binary decomposition.
The transformation $g\left(\cdot\right):\left\{0,1\right\}^{\ell}\rightarrow\left\{0,1\right\}^{\ell}$ induced by this binary code decomposition is defined as follows.
\begin{equation}\label{eq:gBinDefined}
g({\bf u}_0^{\ell-1})={\bf x}_0^{\ell-1} \,\,\,\,\, \text{if  }\, {\bf
x}_0^{\ell-1} \in T_{{\ell}}^{\left({\bf u}_0^{{\ell-1}}\right)}.
\end{equation}
\end{definition}

Following the definition, we can observe, that an SC decision
making on the bits at the input to the polar code encoder (denoted by ${\bf
u}_0^{{\ell-1}}$) given a noisy observation of the output is actually
a decision on the sub-code to which the transmitted vector belongs. As such, deciding on the first bit $u_0$ is equivalent to determining if
the transmitted vector belongs to $T_1^{(0)}$ or to $T_1^{(1)}$.
Once we decided on $u_0$, we assume that we transmitted a codeword
of $T_1^{(u_0)}$ and by deciding on $u_1$ we choose the appropriate
refinement (i.e. sub-code) of $T_1^{(u_0)}$, i.e. we should decide
between the candidates $T_2^{([u_0 \,\,\, 0])}$ and $T_2^{([u_0\,\,\, 1])}$. Due to
this fact, it is not surprising that  the Hamming distance
between the two candidate sub-codes play an important role when
considering the rate of polarization.
\begin{definition}\label{def:partialDistances}
For a binary code decomposition as in Definition \ref{def:
transAssiciatedTheCodes}, the Hamming distances between sub-codes in
the decomposition are defined as follows:

$$D_{min}^{(i)}({\bf u}_0^{i-1})=
\min\left\{d_H\left({\bf c}_0,{\bf c}_1\right)\Big| {\bf c}_0\in
 T_{i+1}^{\left({\bf u}_0^{i-1}\bullet 0\right)},{\bf c}_1\in T_{i+1}^{\left({\bf u}_0^{i-1}\bullet
 1\right)}\right\},$$
$$
D_{min}^{(i)} = \min\left\{D_{min}^{(i)}({\bf u}_0^{i-1})\big|{\bf
u}_0^{i-1}\in \left\{0,1\right\}^{i} \right\},\,\,\,\,\,\,\,i\in[\ell]_{-} .
$$
\normalsize
\end{definition}

A transformation $g\left(\cdot\right)$ can be used as a building block for a recursive construction of a transformation of greater length, in a similar manner to Arikan's method \cite{Arikan}. We specify this construction explicitly in the next definition.
\begin{definition}\label{def:constructG}
Given a transformation $g(\cdot)$ of ${\ell}$ dimensions, we
construct a mapping $g^{(m)}(\cdot)$ of ${\ell}^m$ dimensions  (i.e.
$g^{(m)}(\cdot):\left\{0,1\right\}^{{\ell}^m}\rightarrow\left\{0,1\right\}^{{\ell}^m}$)
in the following recursive fashion.
$$g^{(1)}({\bf u}_0^{\ell-1})=g({\bf u}_0^{\ell-1})\,\,\,;$$

$$g^{(m)}\left({\bf u}_{0}^{\ell^m-1}\right)=\Big[ g^{(m-1)}\left(\left[\gamma_{0,0}\,\,\, \gamma_{1,0}\,\,\, \gamma_{2,0}\,\,\, \ldots\,\,\, \gamma_{{\ell}^{m-1}-1,0}\right]\right) \bullet$$
$$\,\,\,\,\,\,\,\,\,\,\,\,\,\,\,\,\,\,\,\,\,\,\,\,\,g^{(m-1)}\left(\left[\gamma_{0,1}\,\,\, \gamma_{1,1}\,\,\, \gamma_{2,1}\,\,\, \ldots \gamma_{{\ell}^{m-1}-1,1}\right]\right)\bullet \ldots\bullet$$
$$
\,\,\,\,\,\,\,\,\,\,\,\,\,\,\,\,\,\,\,\,\,\,\,\,\,g^{(m-1)}\left(\left[\gamma_{0, {\ell-1}}\,\,\, \gamma_{1, {\ell-1}}\,\,\, \gamma_{2, {\ell-1}}\,\,\, \ldots \,\,\gamma_{{\ell}^{m-1}-1,{\ell-1}}\right]\right)  \Big],
$$
\normalsize
where
$$
\gamma_{i,j}=g_j\left({\bf u}_{i\cdot {\ell}}^{(i+1)\cdot {\ell}-1}\right),
\,\,\,\,\,   i \in \left[{\ell}^{m-1}\right]_{-}, \,\,\,\,\,\,  j \in
\left[{\ell}\right]_{-}.
$$
\end{definition}
The transformation $g^{(m)}(\cdot)$ can be used to encode data and transmit it
over the B-MC channel. Then the method of SC can
be used to decode the information, with decoding complexity of $O\left(2^{\ell}\cdot
N\cdot \log_{\ell}(N)\right)$ (see the appendix for a discussion on the SC decoder).

We use the same definitions and notations for the channel, its corresponding symmetric
capacity and the \bhat parameter, that were used in previous works
\cite{Arikan,Korada,Mori2010}. Note that for uniform binary
random vectors $U_0^{\ell-1}$, and
$X_0^{\ell-1}=g\left(U_0^{\ell-1}\right)$ we have that
$I(Y_0^{\ell-1};U_0^{\ell-1})=I(Y_0^{\ell-1};X_0^{\ell-1})$, because the
transformation $g(\cdot)$ is invertible. Furthermore, since we consider
memoryless channels, we have $I(Y_0^{\ell-1};X_0^{\ell-1}) = {\ell}\cdot
I(Y_0;X_0) = {\ell}\cdot I(\mathcal{W})$, and on the other hand
$$
I(Y_0^{\ell-1};U_0^{\ell-1})  = \sum_{i=0}^{\ell-1}
I(Y_0^{\ell-1};U_i|U_{0}^{i-1}) = \sum_{i=0}^{\ell-1} I(\mathcal{W}^{(i)}).
$$
We define the tree process of the channels generated by the kernels, in
the same way as previous authors did \cite{Arikan,Korada}. A random sequence $\left\{W_n\right\}_{n\geq 0}$ is
defined such that $W_{n} \in \left\{\mathcal{W}^{(i)} \right\}_{i=0}^{{\ell}^n-1}$
with
$$
W_0 = \mathcal{W}
$$
$$
W_{n+1}=W_n^{(B_{n+1})},
$$
where $\left\{B_n\right\}_{n\geq 1}$ is a sequence of i.i.d random variables uniformly distributed over the set $\left[\ell\right]_{-}$. In a similar manner, the symmetric capacity corresponding to the channels  $\left\{I_n\right\}_{n\geq 0} =\left\{I(W_n)\right\}_{n\geq 0} $ and the \bhat parameter random variables  $\left\{Z_n\right\}_{n\geq 0 } =\left\{Z(W_n)\right\}_{n\geq 0 }$ are defined. Just as in \cite[Proposition 8]{Arikan}, it can be shown that the random sequence $\left\{I_n\right\}_{n\geq 0}$ is a bounded martingale, and it is uniformly integrable, which means it converges almost surely to $I_{\infty}$ and that $\mathbb{E}\left\{I_{\infty} \right\} = I(\mathcal{W})$. Now, if we can show that $Z_n \rightarrow Z_{\infty}$ w.h.p such that $Z_{\infty} \in \left\{0,1 \right\}$, by the relations between the channel's symmetric capacity and the \bhat parameter \cite[Proposition 1]{Arikan}, we have that $I_{\infty} \in \left\{0,1\right\}$. But, this means that $\Pr\left(I_{\infty}  = 1\right)=\mathbb{E}\left\{ I_{\infty} \right\} = I(\mathcal{W})$, which is the channel polarization phenomenon.
\begin{proposition}\label{propo: polarizationOfg}
Let $g(\cdot)$ be a binary transformation of ${\ell}$ dimensions,
induced by a binary code decomposition
$\left\{T_0,T_1,...,T_{{\ell}}\right\}$. If  there exists ${\bf
u}_0^{{\ell}-2}\in\left\{0,1\right\}^{{\ell}-1}$  such that
$D_{min}^{(\ell-1)}({\bf u}_0^{{\ell}-2})\geq 2$, then
$\Pr\left(I_{\infty} = 1\right)=I(\mathcal{W})$.
\end{proposition}
\proof Mori and Tanaka \cite[Corollary 11]{Mori2010}
gave sufficient conditions for
\begin{equation}\label{eq:prop1E1}
\lim_{n\rightarrow \infty}\Pr\left(Z_n \in \left(\delta,1-\delta \right) \right) =0,\,\,\,\,\,\forall \delta\in (0,0.5).
\end{equation}
The first condition is that there exists a vector ${\bf u}_0^{{\ell}-2}$, indices
$i,j\in [{\ell}]_{-}$ and permutations $\sigma(\cdot)$, and
$\tau(\cdot)$ on $\left\{0,1\right\}$ such that
$$
g^{({\bf u}_0^{{\ell}-2})}_{i}({  u}_{\ell-1})=\sigma({
u}_{\ell-1})\,\,\,\,\,\,\,\text{and}\,\,\,\,\,\,\, g^{({\bf
u}_0^{{\ell}-2})}_{j}({  u}_{\ell-1})=\tau({  u}_{\ell-1}).
$$
This requirement applies here, because if there exists ${\bf
u}_0^{{\ell}-2}\in\left\{0,1\right\}^{{\ell}-1}$  such that
$D_{min}^{(\ell-1)}({\bf u}_0^{{\ell}-2})\geq 2$, then the two codewords
of the code $T_{\ell-1}^{({\bf u}_0^{{\ell}-2})}$, ${\bf c}_0$ and
${\bf c}_1$, are at  Hamming distance of at least $2$. This means that
there exist  at least two indices $i,j$ such that $c_{0,i}\neq
c_{1,i}$ and $c_{0,j}\neq c_{1,j}$, therefore $g^{({\bf
u}_0^{{\ell}-2})}_{i}({  u}_{\ell-1})$ and $g^{({\bf
u}_0^{{\ell}-2})}_{j}({   u}_{\ell-1})$ are both permutations. The
second condition is that for any ${\bf v}_0^{{\ell}-2}\in\left\{0,1
\right\}^{{\ell}-1}$ there exist an index $m\in[{\ell}]_{-}$ and a permutation
$\mu(\cdot)$ on $\left\{0,1\right\}$ such that
$$
g^{({\bf v}_0^{{\ell}-2})}_{m}({  v}_{\ell-1})=\mu({ v}_{\ell-1}).
$$
This requirement also applies here, because for each ${\bf
v}_0^{{\ell}-2}\in\left\{ 0,1\right\}^{{\ell}-1}$ the two codewords
of the set $T_{\ell-1}^{({\bf v}_0^{{\ell}-2})}$ are at  Hamming
distance of at least 1 apart. This means that (\ref{eq:prop1E1}) holds, which
implies that $I_{\infty} \in \left\{0,1\right\}$ almost surely, and
therefore $\Pr\left(I_{\infty}  = 1\right)=I(\mathcal{W})$. \ \qed

The next proposition on the rate of polarization is an easy consequence of \cite[Theorem 19]{Mori2010} and Proposition \ref{propo: polarizationOfg}.
\begin{proposition}\label{propo:rateOfPolarization}
Let $g(\cdot)$ be a bijective transformation of ${\ell}$ dimensions, induced by
a binary code decomposition $\left\{T_0,T_1,...,T_{{\ell}}\right\}$. If
there exists ${\bf u}_0^{{\ell}-2}\in\left\{0,1\right\}^{{\ell}-1}$
such that $D_{min}^{(\ell-1)}({\bf u}_0^{{\ell}-2})\geq 2$, then

(i) For any $\beta<E(g)$
$$
    \lim_{n\rightarrow \infty} \Pr\left(Z_n\leq 2^{-{\ell}^{n\beta}}\right)=I(\mathcal{W}),
$$

(ii) For any $\beta>E(g)$
    $$
    \lim_{n\rightarrow \infty} \Pr\left(Z_n\geq 2^{-{\ell}^{n\beta}}\right)=1,
    $$
    where
    \begin{equation}\label{eq:exponentEq}
    E(g) = \frac{1}{{\ell}}\sum_{i=0}^{{\ell-1}}\log_{\ell}\left(D_{min}^{(i)}\right).
    \end{equation}
\end{proposition}

\begin{remark} \label{rem:lowerBoundsAndUpperBounds}
Mori and Tanaka \cite{Mori2010} gave two types of exponents: $E_{min}(g)$ and $E_{max}(g)$. The first one is defined as $E_{min}(g)=\frac{1}{\ell}\sum_{i=0}^{\ell-1}\log_{\ell}\left(D_{min}^{(i)}\right)$, and corresponds only to the upper-bound on the \bhat parameter random sequence (as in \textit{(i)} in Proposition \ref{propo:rateOfPolarization}). For the lower-bound on the \bhat parameter random sequence (as in \textit{(ii)} in Proposition \ref{propo:rateOfPolarization}), the exponent is defined as $E_{max}(g)=\frac{1}{\ell}\sum_{i=0}^{\ell-1}\log_{\ell}\left(D_{max}^{(i)}\right)$ (see \cite[Definition 16]{Mori2010} for the appropriate definitions of $D_{min}^{(i)}$ and $D_{max}^{(i)}$ for the non-binary cases).  In Proposition \ref{propo:rateOfPolarization} we used a single type of exponent $E(g)$, because  $E_{max}(g)=E_{min}(g)=E(g)$ if $g(\cdot)$ is a binary kernel or  a non-binary and linear kernel.
\end{remark}
Naturally, we would like to find kernels maximizing $E(g)$. In the next section we consider upper-bounds on the maximum achievable exponent per kernel size $\ell$.
\section{Upper Bounds on the Optimal Exponent}\label{sec:Bounds}
We define the optimal exponent per kernel size $\ell$ as
\begin{equation}\label{eq:El}
E_{\ell}=\max_{g:\{0,1\}^{\ell}\rightarrow\{0,1\}^{\ell}}E\left(g\right).
\end{equation}
Note that Korada \textit{et al.} \cite{Korada} defined $E_{\ell}$  as  the maximization over the set of binary linear kernels, and here we extend the definition for general kernels. Furthermore, a lower-bound on the exponent using the Gilbert-Vershamov technique also applies in this case \cite[Lemma 20]{Korada}. The following lemma is a generalization of \cite[Lemma 18]{Korada}.
\begin{lemma}\label{lem:increasingD}
Let $g:\{0,1\}^{\ell} \rightarrow \{0,1\}^{\ell}$ be a polarizing kernel. Fix $k\in[\ell-1]_{-}$ and define a mapping
\begin{equation}
\tilde{g}\left({\bf v}_0^{\ell-1}\right)=g\left(\left[{\bf v}_0^{k-1} \,\,\,v_{k+1} \,\,\,v_{k}\,\,\, {\bf v}_{k+2}^{\ell-1}\right]\right),
\end{equation}
i.e in this mapping the coordinates $k$ and $k+1$ are swapped. Let $\left\{D_{\text{min}}^{(i)}\right\}_{i=0}^{\ell-1}$ and $\left\{\tilde{D}_{\text{min}}^{(i)}\right\}_{i=0}^{\ell-1}$ denote the partial distance sequences of $g(\cdot)$ and $\tilde{g}(\cdot)$ respectively. If $D_{\text{min}}^{(k)}>D_{\text{min}}^{(k+1)}$ then
\begin{enumerate}[label=(\roman{*})]
  \item $E(g)\leq E(\tilde{g})$
  \item $\tilde{D}_{\text{min}}^{(k)}<\tilde{D}_{\text{min}}^{(k+1)}$
\end{enumerate}
\end{lemma}
\proof
We follow the path of the proof of \cite[Lemma 18]{Korada}.
It will be useful to introduce the following equivalent definition of the partial distance sequence
\begin{equation}\label{eq:partialDistRedef}
D_{\text{min}}^{(i)}=\min\Big\{d_H\left(g\left(\left[{\bf w}_{0}^{i-1}\,\,\,0\,\,\,{\bf u}_{i+1}^{\ell-1}\right]\right),g\left(\left[{\bf w}_{0}^{i-1}\,\,\, 1 \,\,\, {\bf v}_{i+1}^{\ell-1}\right]\right)\right)\Big|{\bf w}_{0}^{i-1},{\bf u}_{i+1}^{\ell-1},{\bf v}_{i+1}^{\ell-1} \Big\}.
\end{equation}
Note that in (\ref{eq:partialDistRedef}), we minimize the Hamming distance $d_H\left(g\left(\left[{\bf w}_{0}^{i-1}\,\,\,0\,\,\,{\bf u}_{i+1}^{\ell-1}\right]\right),g\left(\left[{\bf w}_{0}^{i-1}\,\,\, 1 \,\,\, {\bf v}_{i+1}^{\ell-1}\right]\right)\right)$, over any binary assignment to the vectors ${\bf w}_{0}^{i-1},{\bf u}_{i+1}^{\ell-1}$ and ${\bf v}_{i+1}^{\ell-1}$.
According to this definition, it is easy to see that
\begin{equation}
D_{\text{min}}^{(i)}={\tilde D}_{\text{min}}^{(i)},\,\,\,\,\, i\in [\ell]_{-}\backslash\{k,k+1\}.
\end{equation}
Hence, it suffices to show that
\begin{equation}\label{eq:SuffCond}
D_{\text{min}}^{(k)}\cdot D_{\text{min}}^{(k+1)}\leq {\tilde{D}}_{\text{min}}^{(k)}\cdot {\tilde{D}}_{\text{min}}^{(k+1)}
\end{equation}
in order to prove statement \textit{(i)}.
Using (\ref{eq:partialDistRedef}), we have
\begin{equation}\label{eq:Dk}
  D_{\text{min}}^{(k)}=\min\Big\{d_H\left(g\left(\left[{\bf w}_{0}^{k-1}\,\,\,0\,\,\,{\bf u}_{k+1}^{\ell-1}\right]\right),g\left(\left[{\bf w}_{0}^{k-1}\,\,\,1\,\,\,{\bf v}_{k+1}^{\ell-1}\right)\right)\right]\Big|{\bf w}_{0}^{k-1},{\bf u}_{k+1}^{\ell-1},{\bf v}_{k+1}^{\ell-1} \Big\},
  \end{equation}
  \begin{equation}\label{eq:tDk}
    {\tilde{D}}_{\text{min}}^{(k)}=\min\Big\{d_H\left(g\left(\left[{\bf w}_{0}^{k-1}\,\,\,u_{k+1}\,\,\,0\,\,\,{\bf u}_{k+2}^{\ell-1}\right]\right),g\left(\left[{\bf w}_{0}^{k-1}\,\,\,v_{k+1}\,\,\,1\,\,\,{\bf v}_{k+2}^{\ell-1}\right]\right)\right)\Big|{\bf w}_{0}^{k-1},{\bf u}_{k+1}^{\ell-1},{\bf v}_{k+1}^{\ell-1}\Big\},
    \end{equation}
    \begin{equation}\label{eq:Dkp1}
  D_{\text{min}}^{(k+1)}=\min\Big\{d_H\left(g\left(\left[{\bf w}_{0}^{k}\,\,\,0\,\,\,{\bf u}_{k+2}^{\ell-1}\right]\right),g\left(\left[{\bf w}_{0}^{k}\,\,\,1\,\,\,{\bf v}_{k+2}^{\ell-1}\right]\right)\right)\Big|{\bf w}_{0}^{k},{\bf u}_{k+2}^{\ell-1},{\bf v}_{k+2}^{\ell-1} \Big\},
  \end{equation}
  \begin{equation}\label{eq:tDkp1}
  {\tilde{D}}_{\text{min}}^{(k+1)}=\min\Big\{d_H\left(g\left(\left[{\bf w}_{0}^{k-1}\,\,\,0\,\,\,w_{k}\,\,\,{\bf u}_{k+2}^{\ell-1}\right]\right),g\left(\left[{\bf w}_{0}^{k-1}\,\,\,1\,\,\,w_{k}\,\,\,{\bf v}_{k+2}^{\ell-1}\right]\right)\right)\Big|{\bf w}_{0}^{k},{\bf u}_{k+2}^{\ell-1},{\bf v}_{k+2}^{\ell-1} \Big\}.
\end{equation}
Because the set on which we perform the minimization in (\ref{eq:tDkp1}) is a subset of the set on which we perform the minimization in  (\ref{eq:Dk}) we have that $  D_{\text{min}}^{(k)}\leq {\tilde{D}}_{\text{min}}^{(k+1)} $. On the other hand, the minimization in (\ref{eq:tDk}) can be expressed as $ {\tilde{D}}_{\text{min}}^{(k)}=\min\Big\{\Delta_1,\Delta_2\Big\}$, where
\begin{equation}\label{eq:tDk2a}
\Delta_1 =  \min\Big\{d_H\left(g\left(\left[{\bf w}_{0}^{k-1}\,\,\,u_{k+1}\,\,\,0\,\,\,{\bf u}_{k+2}^{\ell-1}\right]\right),g\left(\left[{\bf w}_{0}^{k-1}\,\,\,u_{k+1}\,\,\,1\,\,\,{\bf v}_{k+2}^{\ell-1}\right]\right)\right)\Big|{\bf w}_{0}^{k-1},{\bf u}_{k+1}^{\ell-1},{\bf v}_{k+2}^{\ell-1} \Big\}
\end{equation}

\begin{equation}\label{eq:tDk2b}
\Delta_2 =  \min\Big\{d_H\left(g\left(\left[{\bf w}_{0}^{k-1}\,\,\,u_{k+1}\,\,\,0\,\,\,{\bf u}_{k+2}^{\ell-1}\right]\right),g\left(\left[{\bf w}_{0}^{k-1}\,\,\,1-u_{k+1}\,\,\,1\,\,\,{\bf v}_{k+2}^{\ell-1}\right]\right)\right)\Big|{\bf w}_{0}^{k-1},{\bf u}_{k+1}^{\ell-1},{\bf v}_{k+2}^{\ell-1} \Big\}.
\end{equation}
We see that $\Delta_1=  D_{\text{min}}^{(k+1)}$ and $\Delta_2\geq D_{\text{min}}^{(k)}$. So,  ${\tilde{D}}_{\text{min}}^{(k)}=D_{\text{min}}^{(k+1)}$, because $D_{\text{min}}^{(k)}>D_{\text{min}}^{(k+1)}$. Therefore this proves (\ref{eq:SuffCond}) and as a consequence it also proves statement \textit{(i)}.
Now,
$$
{\tilde{D}}_{\text{min}}^{(k)}=D_{\text{min}}^{(k+1)}<D_{\text{min}}^{(k)}\leq {\tilde{D}}_{\text{min}}^{(k+1)},
$$
which results in statement \textit{(ii)}. \qed

Lemma \ref{lem:increasingD} implies that when seeking the optimal exponent, $E_{\ell}$, for a given kernel size $\ell$, it suffices to consider kernels with non-decreasing partial distance sequences.
This observation also yields the following lemma.
\begin{lemma}[\cite{Korada},Lemma 22]\label{lem:upperBound}
Let $d(n,k)$ denote the largest possible minimum distance of a binary code of length $n$ and size $2^k$. Then,
\begin{equation}\label{eq:UBKorada1}
E_{\ell}\leq \frac{1}{\ell}\sum_{i=0}^{\ell-1}\log_{\ell}\left(d(\ell,\ell-i)\right).
\end{equation}
\end{lemma}
\proof
Consider a polarizing kernel $g(\cdot)$ of $\ell$ dimensions, that has a partial distance sequence $\left\{D_{\text{min}}^{(i)}\right\}_{i=0}^{\ell-1}$. As a consequence of Lemma \ref{lem:increasingD}, we can assume that the sequence is non-decreasing (otherwise, we can find a kernel that has a non-decreasing sequence with at least the same exponent). Note that
\begin{equation}\label{eq:UBKorada1prf}
D_{\text{min}}^{(k)} = \min_{i\geq k} D_{\text{min}}^{(i)}=\min_{{\bf u}_0^{k-1}}\Big\{\min\big\{d_H({\bf c}_0, {\bf c}_1)\Big | {\bf c}_0,{\bf c}_1\in T_{k}^{\left({\bf u}_0^{k-1}\right)},{\bf c}_0\neq {\bf c}_1 \big\}\Big\}\leq d(\ell,\ell-k),
\end{equation}
where the second inequality is due to the fact that each of the codes in the inner minimum, (i.e. $T_{k}^{\left({\bf u}_0^{k-1}\right)}$), is of size $2^{\ell-k}$ and length $\ell$. \qed

As already noted by Korada \textit{et al.} \cite{Korada}, the shortcoming of (\ref{eq:UBKorada1}) as an upper-bound, is that the dependencies between the partial distances are not exploited. For binary and linear kernels, \cite[Lemma 26]{Korada} gives an improved upper-bound utilizing these dependencies. We now turn to develop an upper-bound  that is applicable to general kernels. The basic idea behind this bound, is to express the partial distance sequence of a kernel, in terms of a distance distribution of a code.

For a code $\mathcal{C}$ of length $\ell$ and size $M$ we define the distance distribution as
\begin{equation}
B_i=\frac{1}{M}\left|\left\{({\bf c}_0,{\bf c}_1)\big| d_H({\bf c}_0,{\bf c}_1)=i \right\} \right|, \,\,\,\,\  i \in [\ell+1]_{-}.
\end{equation}
Note that $B_0 = 1$ and
\begin{equation}\label{eq:Req0}
\sum_{i=1}^{\ell}B_i=M-1.
\end{equation}
Given a non-decreasing partial distance sequence $\left\{D_{\text{min}}^{(i)}\right\}_{i=0}^{\ell-1}$ we choose an arbitrary $k\in [\ell]_{-}$ and consider the sub-sequence $\left\{D_{\text{min}}^{(i)}\right\}_{i=k}^{\ell-1}$. Using the reasoning that led to (\ref{eq:UBKorada1prf}), we  observe that we need to consider the sub-codes $\left\{T_{k}^{\left({\bf u}_0^{k-1}\right)}\right\}_{{\bf u}_0^{k-1}\in\{0,1\}^{k}}$ of size $M=2^{\ell-k}$, however whereas in (\ref{eq:UBKorada1}) we considered only the minimum distance, here we  may have additional constraints on the distance distribution of the code.

Let us begin by understanding the meaning of $D_{\text{min}}^{(\ell-1)}$ (the last element of the sequence). By definition, the code $T_{k}^{\left({\bf u}_0^{k-1}\right)}$ is decomposed into $\frac{2^{\ell-k}}{2}$ sub-codes of size $2$, such that in each one the distance between the two codewords is at least $D_{\text{min}}^{(\ell-1)}$. This means that we must satisfy the following requirement
\begin{equation}\label{eq:Req1}
\sum_{i=D_{\text{min}}^{(\ell-1)}}^{\ell} B_i\geq 1,
\end{equation}
where $\left\{B_i\right\}_{i=0}^{\ell}$ is the distance distribution  of  $T_{k}^{\left({\bf u}_0^{k-1}\right)}$.

Now, let us proceed to $D_{\text{min}}^{(\ell-2)}$. This item implies that there are $\frac{2^{\ell-k}}{2^2}$ sub-codes of $T_{k}^{\left({\bf u}_0^{k-1}\right)}$ of four codewords that each one of them can be decomposed into two sub-codes with  Hamming distance of at least $D_{\text{min}}^{(\ell-2)}$. From this, we deduce that there are  $2\cdot 2^{\ell-k}$ pairs of codewords having their distance of at least $D_{\text{min}}^{(\ell-2)}$. These pairs are in addition to the the ones we counted in (\ref{eq:Req1}). Thus, because we assume that the partial distance sequence is non-decreasing, we have the following requirement
\begin{equation}\label{eq:Req2}
\sum_{i=D_{\text{min}}^{(\ell-2)}}^{\ell} B_i\geq 3.
\end{equation}
Note that if $D_{\text{min}}^{(\ell-2)}=D_{\text{min}}^{(\ell-1)}$ then (\ref{eq:Req1}) is redundant given (\ref{eq:Req2}).

In the general case, when considering $D_{\text{min}}^{(\ell-r)}$, where $r \in [\ell-k]$, we need to take into account the
$\frac{2^{\ell-k}}{2^{r}}$ sub-codes of $T_{k}^{\left({\bf u}_0^{k-1}\right)}$, each one of size $2^{r}$ and each one can be partitioned into two sub-codes with Hamming distance of at least $D_{\text{min}}^{(\ell-r)}$. So, there are
$
2\cdot \frac{2^{\ell-k}}{2^{r}}\cdot \left(2^{r}\right)^2 = M \cdot 2^r
$
  pairs of codewords (that were not counted at the previous steps)  such that their distance is at least $D_{\text{min}}^{(\ell-r)}$. Summarizing, we get the following set of $\ell-k$ inequalities
\begin{equation}\label{eq:ReqGen}
 \sum_{i=D_{\text{min}}^{(\ell-r)}}^{\ell} B_i\geq \sum_{j=0}^{r-1}2^j = 2^{r}-1,  \,\,\,\,\,\,\,\,\,\,\,\,\,  r\in [\ell-k].
\end{equation}
By Delsarte \cite{Delsarte73}, the following linear inequalities on the distance distribution are valid.
\begin{equation}\label{eq:ReqDelsarte}
\sum_{j=1}^{\ell}B_j\cdot P_i(j) \geq -{\ell  \choose i}, \,\,\,\,\,\,\,\,\,  i \in [\ell+1]_{-},
\end{equation}
where $P_k(x)$ is the Krawtchouk polynomial, which is defined as
\begin{equation}\label{eq:Kraw}
P_k(x)=\sum_{m=0}^{k}(-1)^m{x \choose m}{\ell - x \choose k-m}.
\end{equation}
In addition, the following is also an obvious requirement
\begin{equation}\label{eq:ReqNonNegative}
B_i \geq 0 \,\,\,\,\,\,\, i\in [\ell].
\end{equation}
We see that  requirements (\ref{eq:Req0}),(\ref{eq:ReqGen}),(\ref{eq:ReqDelsarte}) and (\ref{eq:ReqNonNegative}) are all linear. A partial distance sequence that corresponds to a kernel must be able to  satisfy these constraints for every $k\in [\ell]$. So, taking the maximum exponent corresponding to a partial distance sequence that fulfills the requirements for each $k\in[\ell]$ results in an upper-bound on the exponent.  Checking the validity of a sequence can be done by linear programming methods (we need to check if the polytope is not empty). We now turn to give two simple examples of the method, and then we present a variation on this development that leads to a stronger bound.
\begin{example}
Consider $\ell = 3$. Let $\left\{D_{\text{min}}^{(i)}\right\}_{i=0}^{2}$ be the partial distance sequence of the optimal exponent of size $3$. Note first that by the Singleton bound $D_{\text{min}}^{(k)}\leq k+1$. We first consider the possibility that $D_{\text{min}}^{(2)}=3$ and $D_{\text{min}}^{(1)}=2$. This assumption is translated by (\ref{eq:Req0}) and (\ref{eq:ReqGen}) to
\begin{equation}\label{eq:exl3_1}
B_2+B_3=3\,\,\,\,\,\,\,\,\,\,, B_3\geq 1 \,\,\,\,\,,\,\,\,\, B_2\geq 0
\end{equation}

\begin{equation}\label{eq:exl3_2}
-B_2-3\cdot B_3 \geq -3\,\,\,\, \Longrightarrow_{B_3\geq 1,B_2\geq 0}  B_2 =0,B_3 = 1
\end{equation}
and this is a contradiction to (\ref{eq:exl3_1}).
The next best candidate is a sequence having $D_{\text{min}}^{(1)}=D_{\text{min}}^{(2)}=2$. This sequence can be achieved by a binary linear kernel induced by the generating matrix
$$
\left[
  \begin{array}{ccc}
    1 & 0 & 0 \\
    1 & 1 & 0 \\
    0 & 1 & 1 \\
  \end{array}
\right].
$$
This proves that $E_{3}=\frac{1}{3}\log_3{4}\approx0.42062$.
\end{example}

\begin{example}
Consider $\ell = 4$. Let $\left\{D_{\text{min}}^{(i)}\right\}_{i=0}^{3}$ be the partial distance sequence of the optimal exponent of size $4$. We first consider the possibility that $D_{\text{min}}^{(3)}=D_{\text{min}}^{(2)}=3$  (if this possibility is eliminated it means that $D_{\text{min}}^{(3)}=4,D_{\text{min}}^{(2)}=3$ is also not possible).
  Conditions (\ref{eq:Req0}) and (\ref{eq:ReqGen}) are translated to
\begin{equation}\label{eq:exl4_1}
B_3+B_4=3\,\,\,\,\, \,\,\,\, B_3,B_4\geq 0
\end{equation}
By (\ref{eq:ReqDelsarte}) for $i=1$ we have
$$
B_3\cdot P_1(3)+B_4\cdot P_1(4) \geq -4
$$
$$
-2\cdot B_3-4\cdot B_4 \geq -4\Longrightarrow_{\text{(\ref{eq:exl4_1})}}B_3 +2(3-B_3)\leq 2\Longrightarrow B_3\geq 4
$$
which is a contradiction to (\ref{eq:exl4_1}).
The next best candidate is $$D_{\text{min}}^{(3)}=4,D_{\text{min}}^{(2)}=2,D_{\text{min}}^{(1)}=2,D_{\text{min}}^{(0)}=1,$$
which can be achieved by a binary linear kernel induced by the generating matrix
$$
\left[
  \begin{array}{cc}
    1 & 0 \\
    1 & 1 \\
  \end{array}
\right]^{\otimes2}.
$$
This proves that $E_4=0.5$.
\end{example}

The notion of translating the partial distance sequence into requirements on distance distributions can be further refined. This approach leads to a better bound that we now turn to present.
As we did before, we begin our discussion  by considering the sub-sequence $\left\{D_{\text{min}}^{(i)}\right\}_{i=k}^{\ell-1}$. We start by giving an interpretation to $D_{\text{min}}^{(\ell-1)}$ (the last element of the sequence). By definition, the code $T_{k}^{\left({\bf u}_0^{k-1}\right)}$ is decomposed into $\frac{2^{\ell-k}}{2}$ sub-codes of size $2$, where in each one the distances between the two codewords are at least $D_{\text{min}}^{(\ell-1)}$. Denote by $B_{i}^{\left({\bf u}_0^{\ell-2}\right)} \,\,\,\, i\in [\ell]$ the partial distance distribution of the sub-code $T_{\ell-1}^{\left({\bf u}_0^{\ell-2}\right)}$ of the code $T_{k}^{\left({\bf u}_0^{k-1}\right)}$. By definition we have
\begin{equation}
B_{i}^{\left({\bf u}_0^{\ell-2}\right)} = \frac{1}{2}\left|\left\{d_H({\bf c}_0,{\bf c}_1 )=i\Big| {\bf c}_0,{\bf c}_1\in T_{\ell-1}^{\left({\bf u}_0^{\ell-2}\right)}\right\}\right|.
\end{equation}
This leads to
\begin{equation}
\sum_{i=D_{\text{min}}^{(\ell-1)}}^{\ell} B_{i}^{\left({\bf u}_0^{\ell-2}\right)} = 1, \,\,\,\,\, \forall {\bf u}_0^{\ell-2}\in\{0,1\}^{\ell-1},
\end{equation}
\begin{equation}
\sum_{j=1}^{\ell}B_{j}^{\left({\bf u}_0^{\ell-2}\right)}\cdot P_i(j) \geq -{\ell  \choose i}, \,\,\,\,\,\,\,\,\, i \in [\ell+1]_{-},\forall {\bf u}_0^{\ell-2}\in\{0,1\}^{\ell-1}.
\end{equation}
Denote by $\bar{B}_{i}^{(\ell-1)}$  the average of these distributions over all the sub-codes of $T_{k}^{\left({\bf u}_0^{k-1}\right)}$, i.e.
\begin{equation}
\bar{B}_{i}^{(\ell-1)}=\frac{1}{2^{\ell-k-1}}\sum_{{\bf u}_k^{\ell-2}\in\{0,1\}^{\ell-k-1}}B_{i}^{\left({\bf u}_0^{\ell-2}\right)}, \,\,\,\,\,\,\,\,\,\, i\in[\ell].
\end{equation}
Note that
\begin{equation}
\bar{B}_{i}^{(\ell-1)} = \frac{1}{M}\left|\left\{d_H({\bf c}_0,{\bf c}_1 )=i\Big| {\bf c}_0,{\bf c}_1\in T_{k}^{\left({\bf u}_0^{\ell-2}\right)}, {\bf u}_k^{\ell-2}\in\{0,1\}^{\ell-1-k}\right\}\right|
\end{equation}
and
\begin{equation}
\sum_{i=D_{\text{min}}^{(\ell-1)}}^{\ell}\bar{B}_{i}^{(\ell-1)} = 1 ,
\end{equation}
\begin{equation}
\sum_{j=1}^{\ell}\bar{B}_{j}^{(\ell-1)}\cdot P_i(j) \geq -{\ell  \choose i}, \,\,\,\,\,\,\,\,\,  i \in [\ell+1]_{-}.
\end{equation}

Let us proceed to $D_{\text{min}}^{(\ell-2)}$.
By definition, the code $T_{k}^{\left({\bf u}_0^{k-1}\right)}$ is decomposed into $\frac{2^{\ell-k}}{4}$ sub-codes of size $4$, where in each one the distance between the two codewords is at least $D_{\text{min}}^{(\ell-2)}$. Denote by $B_{i}^{\left({\bf u}_0^{\ell-3}\right)} \,\,\,\, i\in [\ell]$, the  distance distribution of the sub-code $T_{\ell-2}^{\left({\bf u}_0^{\ell-3}\right)}$ of the code $T_{k}^{\left({\bf u}_0^{k-1}\right)}$.
\begin{equation}
B_{i}^{\left({\bf u}_0^{\ell-3}\right)} = \frac{1}{4}\left|\left\{d_H({\bf c}_0,{\bf c}_1 )=i\Big| {\bf c}_0,{\bf c}_1\in T_{\ell-2}^{\left({\bf u}_0^{\ell-3}\right)}\right\}\right|.
\end{equation}
Note that
\begin{equation}
B_{i}^{\left({\bf u}_0^{\ell-3}\right)}\geq \frac{1}{2}\left(B_{i}^{\left({\bf u}_0^{\ell-3}\bullet 0\right)}+B_{i}^{\left({\bf u}_0^{\ell-3}\bullet 1\right)}\right).
\end{equation}
So by introducing the average distance distribution
\begin{equation}
\bar{B}_{i}^{(\ell-2)}=\frac{1}{2^{\ell-k-2}}\sum_{{\bf u}_k^{\ell-3}\in\{0,1\}^{\ell-k-2}}B_{i}^{\left({\bf u}_0^{\ell-3}\right)}, \,\,\,\,\,\,\,\, i\in[\ell],
\end{equation}
we get
\begin{equation}
\sum_{i=D_{\text{min}}^{(\ell-2)}}^{\ell}\bar{B}_{i}^{(\ell-2)} = 3 ,
\end{equation}
\begin{equation}
\sum_{j=1}^{\ell}\bar{B}_{j}^{(\ell-2)}\cdot P_i(j) \geq -{\ell  \choose i}, \,\,\,\,\,\,\,\,\,  i \in [\ell+1]_{-}.
\end{equation}
and
\begin{equation}
\bar{B}_{i}^{(\ell-2)} - \bar{B}_{i}^{(\ell-1)}\geq 0, \,\,\,\,\,\,\,\,\,  i \in [\ell].
\end{equation}
In the general case, when taking $D_{\text{min}}^{(\ell-r)}$ into account, where $1 \leq r \leq \ell-k$, we essentially consider the $\frac{2^{\ell-k}}{2^{r}}$ sub-codes of $T_{k}^{\left({\bf u}_0^{k-1}\right)}$, each one of size $2^{r}$ and each one can be partitioned into two sub-codes of size $2^r$ with Hamming distance of at least $D_{\text{min}}^{(\ell-r)}$ between them. Denote the distance distribution of the sub-code $T_{\ell -r}^{\left({\bf u}_0^{\ell -r-1}\right)}$ as  $\left\{B_{i}^{\left({\bf u}_0^{\ell-r-1}\right)}\right\}_{i\in[\ell]}$ and the average distance distribution  as $\left\{\bar{B}_{i}^{(\ell-r)}\right\}_{i\in[\ell]}$. We have
\begin{equation}
B_{i}^{\left({\bf u}_0^{\ell-r-1)}\right)} = \frac{1}{2^{r}}\left|\left\{d_H({\bf c}_0,{\bf c}_1 )=i\Big| {\bf c}_0,{\bf c}_1\in T_{\ell-r}^{\left({\bf u}_0^{\ell-r-1}\right)}\right\}\right|,
\end{equation}
\begin{equation}
\bar{B}_{i}^{(\ell-r)}=\frac{1}{2^{\ell-k-r}}\sum_{{\bf u}_k^{\ell-r-1}\in\{0,1\}^{\ell-k-r}}B_{i}^{\left({\bf u}_0^{\ell-r-1}\right)}, \,\,\,\,\,\,\,\,\, i\in[\ell],
\end{equation}

which results in
\begin{equation}
 \sum_{i=D_{\text{min}}^{(\ell-r)}}^{\ell}\bar{B}_{i}^{(\ell-r)}= \sum_{j=0}^{r-1}2^j = 2^{r}-1,
\end{equation}

\begin{equation}
\bar{B}_{i}^{(\ell-r)}-\bar{B}_{i}^{(\ell-r+1)}\geq 0,\,\,\,\,\,\,\,\,\,  i \in[\ell],
\end{equation}
\begin{equation}
\sum_{j=1}^{\ell}\bar{B}_{j}^{(\ell-r)}\cdot P_i(j) \geq -{\ell  \choose i}, \,\,\,\,\,\,\,\,\,  i \in[\ell+1]_{-}.
\end{equation}
We are now ready to summarize this development.

\begin{definition}\label{def:LPValid}
Let $\left\{ D_i \right\}_{i=0}^{\ell-1}$ be a monotone non-decreasing sequence of non-negative integral numbers, such that $ D_i \leq d(\ell,\ell-i)$.
We say that this sequence is an $\ell$-dimensions \textit{Linear Programming (LP) valid} sequence if the polytope defined by (\ref{eq:poly1}) - (\ref{eq:poly3}) on the non-negative variables $\left\{\bar{B}_i^{(k)}\big|  D_{k} \leq i \leq \ell  ,\,\,\,\,\,k\in[\ell]_{-} \right\} $ is not empty:
\begin{equation}\label{eq:poly1}
 \sum_{i=D_{\ell-r}}^{\ell}\bar{B}_{i}^{(\ell-r)}= \sum_{i=0}^{r-1}2^i = 2^{r}-1,\,\,\,\,\,\,  r\in [\ell],
\end{equation}

\begin{equation}\label{eq:poly2}
\bar{B}_{i}^{(\ell-r)}-\bar{B}_{i}^{(\ell-r+1)}\geq 0,\,\,\,\,\,D_{\ell-r+1} \leq i \leq \ell,\,\,\,\,\,\,\, r \in [\ell-1],
\end{equation}
\begin{equation}\label{eq:poly3}
\sum_{j=D_{\ell-r}}^{\ell}\bar{B}_{j}^{(\ell-r)}\cdot P_i(j) \geq -{\ell  \choose i}, \,\,\,\,\,\,\,\,\, i \in [\ell+1]_{-},\,\,\,\,\,\,\,\, r\in [\ell].
\end{equation}
\end{definition}

\begin{proposition}\label{propo:LPvalid}
If  $\left\{ D_{\text{min}}^{(i)}\right\}_{i=0}^{\ell-1}$ is a partial distance sequence corresponding to some binary $\ell$ dimensions kernel $g(\cdot)$, then $\left\{ D_{\text{min}}^{(i)}\right\}_{i=0}^{\ell-1}$  is an $\ell$-dimensions LP-valid sequence.
\end{proposition}

We denote by $\mathcal{V}^{(\ell)}_{\text{LP}}$ the set of all the $\ell$-dimensions $LP$-valid sequences. The following proposition is an easy consequence of Proposition \ref{propo:LPvalid}.
\begin{proposition}\label{propo:UBLP}
\begin{equation}\label{eq:UBOur}
E_{\ell}\leq \max_{\left\{D_k \right\}_{k\in[\ell]_{-}} \in \mathcal{V}^{(\ell)}_{\text{LP}}} \frac{1}{\ell}\sum_{i=0}^{\ell-1} \log_{\ell}D_i.
\end{equation}
\end{proposition}
The method of Proposition \ref{propo:UBLP} can be easily generalized to non-binary kernels with alphabet size $q$, by applying the following changes to Definition \ref{def:LPValid}.
\begin{itemize}
\item In (\ref{eq:poly1}), the right-hand side of the equation is  replaced by $q^{r}-1$.
\item In (\ref{eq:poly3}), the Krawtchouk polynomial, is replaced by its non-binary version
    \begin{equation}\label{eq:KrawGen}
P_k(x)=\sum_{m=0}^{k}(-1)^m(q-1)^{k-m}{x \choose m}{\ell - x \choose k-m}.
\end{equation}
The right-hand side of  (\ref{eq:poly3})  is multiplied by $(q-1)^i$.
\end{itemize}
This leads to an upper-bound  on Mori and Tanaka's exponent $E_{min}(g)$   \cite[Theorem 19]{Mori2010}. Note that the distinction between $E_{min}(g)$ and $E(g)$ is required here because the kernel $g(\cdot)$ is non-binary (see Remark \ref{rem:lowerBoundsAndUpperBounds} for further details).

We computed the bound for several instances of $\ell$ by carefully enumerating the sequences in $\mathcal{V}^{(\ell)}_{\text{LP}}$ using Wolfram \textit{Mathematica} LP-Solver. The enumeration process involves generating a linear program having a polytope that is defined by the partial sequence, and using the LP solver for solving it. If the solver could not find a solution, then in our case it means that the polytope is infeasible (this is because the polytope is always bounded). For $\ell\leq 17$ we used both the \textit{Mathematica} Simplex algorithm implementation  with infinite precision and also the Interior-Point algorithm with finite precision and received the same results. Due to the long running time of the enumeration algorithm, we had to retreat to the interior point algorithm for $\ell>17$. For these cases, because of the finite computer precision of the software implementation and the limited number of iterations of the algorithm these results might be inaccurate.

Table \ref{tbl:UBEl} contains the results of the enumeration for $5\leq\ell\leq 25$. The table contains the upper-bounds $E_{\ell}$, and the  \textit{LP valid} sequences that correspond to these bounds. For comparison, we also generated upper-bounds based on Lemma \ref{lem:upperBound} and Argell's table of upper-bounds for unrestricted binary codes  \cite{bndsForUnrestrictedCodes}.    We note that the   \textit{LP valid} sequences   are not necessarily achievable by a kernel with corresponding size $\ell$.

Table \ref{tbl:UBE2} contains examples of the upper-bound for non-binary kernels (these results were obtained using the Simplex algorithm with infinite precision). By \textit{Construction X} of Sloane \textit{et al.} \cite{Sloane1972}, if there exists a chain $(n,k,d_0)-(n,k-1,d_1)$, then there exists a code $\left(n+1,k,\min\left\{d_0+1,d_1\right\}\right)$.  Therefore, if there does not exist a code with the latter parameters, then the chain is also invalid. We used this idea, for the entry of $q=4$ and $\ell=5$, to eliminate sequences with prefix $(1,2,3,\ldots)$, because this implies an existence of a quaternary   $(n=6,M=2^4,d=3)$ code, which contradicts the Hamming bound. The Hamming bound also forbids the  same prefix for $q=8$ and $\ell=9$.

In the
next section, we give examples of good kernels, that are derived by utilizing results about known code decompositions, for $14\leq\ell\leq16$ that achieve the optimal exponent.
\begin{table}
\center
\begin{tabular}{|c|l|c|c|}
   \hline
 ${\ell}$& \textit{LP Valid} optimal sequence & Upper-bound  on $E_{\ell}$ & Upper-bound on $E_{\ell}$\\
& & (Proposition \ref{propo:UBLP}) & (Lemma \ref{lem:upperBound}) \\
  \hline
  5 & $1, 2, 2, 2, 4$ & $0.43067$ & $0.50879$\\
  6 & $1, 2, 2, 2, 4, 4$ & $0.45132$&$0.52676 $ \\
  7 & $ 1,2,2,2,4,4,4$& $0.45798$&$0.52883$\\
  8 & $1,2,2,2,4,4,4,8$&$0.5$&$0.51341$\\
  9 & $1,2,2,2,2,4,4,6,6$ & $0.46162$&$0.50263$\\
  10 &$1,2,2,2,2,4,4,4,6,8$&$0.46915$&$0.50614$\\
  11& $1,2,2,2,2,4,4,4,6,6,8$&$0.47748$& $0.51923 $\\
   12  & $1, 2, 2, 2, 2, 4, 4, 4, 6, 6, 6, 12$  & $0.49605$&$0.52677$ \\
   13 & $1, 2, 2, 2, 2, 4, 4, 4, 6, 6, 6, 8, 10$ & $0.50049$&$0.53184$ \\
   14 & $1, 2, 2, 2, 2, 4, 4, 4, 6, 6, 6, 8, 8, 8$ & 0.50194&$0.54146$ \\
   15 & $1, 2, 2, 2, 2, 4, 4, 4, 6, 6, 6, 8, 8, 8, 8$ & $0.50773$&$0.54797$ \\
   16 & $1, 2, 2, 2, 2, 4, 4, 4, 6, 6, 6, 8, 8, 8, 8, 16$ & $0.52742$&$0.53245$ \\
   17 & $1, 1, 2, 2, 2, 3, 4, 4, 5, 6, 6, 7, 8, 8, 8, 9, 16$ & $0.50447$&$0.52673$ \\
   18 & $1, 1, 2, 2, 2, 3, 4, 4, 5, 6, 6, 7, 8, 8, 8, 9, 12, 12$ & $0.50925$&$0.53466$ \\
   19 & $1, 1, 2, 2, 2, 3, 4, 4, 5, 6, 6, 7, 8, 8, 8, 9, 10, 12, 12$ & $0.51475$&$0.53934$\\
  20 & $1, 1, 2, 2, 2, 3, 4, 4, 5, 6, 6, 7, 8, 8, 8, 8, 10, 10, 12, 16$& $0.52190$&$0.54385$\\
  21& $1, 2, 2, 2, 2, 2, 4, 4, 4, 6, 6, 6, 8, 8, 8, 8, 10, 10, 10, 12, 18$& $0.52554$&$0.54381$\\
   22& $1, 1, 2, 2, 2, 3, 4, 4, 4, 5, 6, 6, 7, 8, 8, 9, 10, 10, 11, 12, 12, 16$ &$0.52317$&$0.54454$\\
   23& $1, 1, 2, 2, 2, 3, 4, 4, 4, 5, 6, 6, 7, 8, 8, 9, 10, 10, 10, 11, 12, 14, 16$& $0.52739$&$0.54788$\\
24 & $1, 1, 2, 2, 2, 3, 4, 4, 4, 5, 6, 6, 7, 8, 8, 8, 9, 10, 11, 12, 12, 12, 14, 20$& $0.53362$&$0.54840$\\
   25 & $1, 1, 2, 2, 2, 3, 4, 4, 4, 5, 6, 6, 7, 8, 8, 8, 9, 10, 10, 12, 12, 12, 12, 15, 20$& $0.53633$&$0.54935$\\
  \hline
\end{tabular}
 \normalsize
\caption{Upper-bounds on $E_{\ell}$ computed by Wolfram \textit{Mathematica} LP-Solver according to Proposition \ref{propo:UBLP} ($\ell$ is the size of the kernel). The optimal sequence is the sequence that corresponds to the exponent of Proposition \ref{propo:UBLP}. For comparison, upper-bounds based on Lemma \ref{lem:upperBound} are listed in the rightmost column. }\label{tbl:UBEl}
\end{table}

\begin{table}
\center
\begin{tabular}{|c|c|l|c|}
   \hline
 $q$ &${\ell}$& \textit{LP Valid} optimal sequence & Upper-bound  on $E_{\ell}$  \\
  \hline
  $4$ & $\leq 4$ & $1, 2, ...,\ell$ &
  $\frac{1}{\ell}\sum_{i=1}^{\ell}\log_{\ell}(i)$\\
  $4$ & $5$ & $1,2,2,4,4$ & $0.51681$\\
  $4$ & $6$ & $1,2,2,4,4,6$ & $0.55351$\\
  $4$ & $7$ & $1,2,2,3,4,5,7$ & $0.54521$\\
  $4$ & $8$ & $1,2,2,3,4,5,6,8$ & $0.56216$\\
  $4$ & $16$ & $1,2,2,3,4,4,5,6,7,8,9,10,10,12,12,16$ & $0.61379$\\
  $8$ & $\leq 8$ & $1, 2, ...,\ell$ &
  $\frac{1}{\ell}\sum_{i=1}^{\ell}\log_{\ell}(i)$\\
  $8$ & $9$ & $1,2,2,4,5,6,7,8,8$ & $0.62091$\\
  $8$ & $10$ & $1,2,2,4,5,6,7,8,8,10$ & $0.63325$\\
  $8$ & $11$ & $1,2,2,3,5,6,7,8,8,8,11$ & $0.62434$\\
  $8$ & $16$ & $1,2,2,3,4,5,6,7,8,9,10,11,12,13,14,14$ & $0.64297$\\
  \hline
\end{tabular}
 \normalsize
\caption{Upper-bounds on non-binary $E_{\ell}$ computed by Wolfram \textit{Mathematica} LP-Solver according to Proposition \ref{propo:UBLP} ($q$ is the alphabet size and $\ell$ is  the size of the kernel). The optimal sequence it the sequence that corresponds to the exponent on the right column. }\label{tbl:UBE2}
\end{table}

\section{Designing Kernels by  Known Code Decompositions}\label{sec:ExmpOfGoodKernels}
As we noticed in  Section \ref{sec:CodesDecom}, the  exponent, $E(g)$, is influenced by Hamming distances between the sub-codes in the binary partition $\left\{T_0,...,T_{\ell}\right\}$. In this section, we use a particular method for deriving good partial distance sequences by using known decompositions, which are not necessarily binary decompositions. The following observation links between general decompositions and  binary decompositions.

\begin{observation}\label{obs:linkBetSetPartCosetDecomo}
If there exists a code decomposition of $\left\{0,1\right\}^{\ell}$ with the following chain of parameters
$$
({\ell},k_0,d_0)-({\ell},k_1,d_1)-...-({\ell},k_{m-1},d_{m-1}),
$$
then there exists a binary code decomposition of $\left\{0,1\right\}^{\ell}$, such that
$$
D_{min}^{(i)} \geq d_{j}\,\,\,\,\, \text{where} \,\,\,\, \ell-k_{j}\leq
i \leq \ell-1-k_{j+1},\,\,\,\,\,\,\,
$$
$$
j\in[m]_{-},\,\,i\in[{\ell}-1],\,\,
k_{m}=0.
$$
\end{observation}

The next statement  is an easy corollary that follows from (\ref{eq:exponentEq}) and the previous observation.
\begin{corollary}\label{obs:linkBetSetPartCosetAndErrorExponent}
If there exists a code decomposition of $\left\{0,1\right\}^{\ell}$
with the following chain of parameters
$$
({\ell},k_0,d_0)-({\ell},k_1,d_1)-...-({\ell},k_{m-1},d_{m-1}),
$$
then there exists a binary kernel $g(\cdot)$  of ${\ell}$ dimensions
induced by a binary code decomposition
$\left\{T_0,...,T_{{\ell}}\right\}$ such that
\begin{equation}\label{eq:partCosetsE1LB}
E(g) \geq (1/{\ell})\cdot \sum_{j=0}^{m-1} (k_j-k_{j+1})\cdot \log_{\ell}\left(d_j\right),
\end{equation}
where $k_{m}=0$.
\end{corollary}

A list of code  decompositions for ${\ell}\leq 16$ was given in \cite[Table 5]{LitsynTblBinaryCodes}.
Using this list, Corollary \ref{obs:linkBetSetPartCosetAndErrorExponent} and Propositions \ref{propo: polarizationOfg} and \ref{propo:rateOfPolarization}, we can construct polarizing non-linear kernels and obtain lower-bounds on their  exponent $E(g)$.   Table \ref{tbl:errorExp} contains a list of code decompositions that give lower-bounds on $E(g)$ that are greater than $0.5$. At the chain
description column of the table, the code length equals ${\ell}$ for all the sub-codes,
and was omitted from the chain for brevity. Note that the second entry of the table contains the same exponent as the kernel constructed by Korada \textit{et al.} \cite{Korada}. It was proven that this is the best linear binary kernel of size $16$, and that all the binary linear kernels of size $<15$ have exponents $\leq 0.5$. The first entry of the table gives a non-linear decomposition resulting in a non-linear kernel having a better exponent. In fact, this exponent is even better than  all the exponents that appeared in \cite[Table 1]{Korada}. Furthermore, entries $1,3$ and $4$ achieve the upper-bound on the exponent per their kernel size as Table \ref{tbl:UBEl} indicates. Thus, the exponent values indicated in Table \ref{tbl:errorExp} are not just lower-bounds, but rather the true exponents. Note that all the upper-bounds on the exponent, corresponding to $\ell\in [16] \backslash\{12,13\}$ in Table \ref{tbl:UBEl}, can be achieved by decompositions from \cite[Table 5]{LitsynTblBinaryCodes}.
The appendix contains details about the decompositions in Table \ref{tbl:errorExp}.
\begin{table}
\center
\begin{tabular}{|c|c|l|c|}
  \hline
  $\#$ &${\ell}$&Chain description&Lower-  \\
     & & &  bound on\\
     & & &  $E(g)$\\
  \hline
  1 & 16  & $(16,1)-(15,2)-(11,4)-(8,6)-(5,8)-(1,16)$  & 0.52742 \\
  2 & 16 & $(16,1)-(15,2)-(11,4)-(7,6)-(5,8)-(1,16)$ & 0.51828 \\
  3 & 15 & $(15,1)-(14,2)-(10,4)-(7,6)-(4,8)$ & 0.50773 \\
  4 & 14 & $(14,1)-(13,2)-(9,4)-(6,6)-(3,8)$ & 0.50194 \\
  \hline
\end{tabular}
 \normalsize
\caption{Code decompositions from \cite[Table
5]{LitsynTblBinaryCodes} with their corresponding lower-bounds on
kernel exponents for the kernels induced by them. }\label{tbl:errorExp}
\end{table}

\section{Designing Kernels by Modifying Known Constructions}\label{sec:ExmpofModifiying}
\begin{table}
\center
\begin{tabular}{|c|l|c|}
   \hline
 ${\ell}$& Partial distance sequence & Lower-bound  on $E_{\ell}$  \\
  \hline
 18 & $1,2,2,2,2,2,4,4,4,4,6,6,8,8,8,8,8,16$ & $0.49521$ \\
   19 & $1,2,2,2,2,2,4,4,4,4,4,6,8,8,8,8,8,8,16$ & $0.49045$\\
  21& $1,2,2,2,2,2,4,4,4,4,4,4,8,8,8,8,8,8,12,12, 12$& $0.49604$\\
   22& $1,2,2,2,2,2,4,4,4,4,4,4,8,8,8,8,8,8,8,12,12,16$ &$0.50118$\\
   23& $1,2,2,2,2,2,4,4,4,4,4,4,8,8,8,8,8,8,8,12,12,12,16$& $0.50705$\\
24 & $1,2,2,2,2,2,4,4,4,4,4,4,8,8,8,8,8,8,8,12,12,12,16,16$& $0.51577$\\
   25 & $1,2,2,2,2,2,4,4,4,4,4,4,4,8,8,8,8,8,8,8,12,12,12,16,16$& $0.50608$\\
  \hline
\end{tabular}
 \normalsize
\caption{Lower-bounds on $E_{\ell}$ derived by modifications of known code constructions in Section \ref{sec:ExmpOfGoodKernels}. }\label{tbl:LBTbl}
\end{table}

In this section, we use modifications such as shortening, puncturing and extending on known code structures to design kernels, having the largest exponents known so far per their kernel size. This will lead to lower-bounds on $E_{\ell}$, which are summarized in Table \ref{tbl:LBTbl}. We begin by recalling the shortening technique in the context of linear polar codes, that was introduced by Korada \textit{et al.} \cite[Section VI]{Korada}. For completeness,  we cite the following statement.
\begin{lemma}[\cite{Korada},Lemma 30]\label{lem:Shortn}
Let $G$ be an $\ell\times\ell$ binary matrix corresponding to a linear kernel $g(\cdot)$ of $\ell$ dimensions, such that $g({\bf u})={\bf u}\cdot G$. Assume that $g(\cdot)$ has a monotone non-decreasing partial distance sequence $\left\{D_{\text{min}}^{(i)}\right\}_{i=0}^{\ell-1}$. If column $j$ of $G$ has its last $'1'$ in row $k\in[\ell]_{-}$, then an $(\ell-1)\times(\ell-1)$ matrix $\tilde{G}$ obtained by adding row $k$ to all the rows having $1$ in column $j$, and then deleting row $k$  and column $j$, induces a linear transformation $\tilde{g}({{\bf v} })={{\bf v}}\cdot {\tilde{G}}$ of $\ell-1$ dimensions, with partial distance distribution $\left\{\tilde{D}_{\text{min}}^{(i)}\right\}_{i=0}^{\ell-2}$
such that
\begin{equation}
\tilde{D}_{\text{min}}^{(i)} \geq D_{\text{min}}^{(i)}\,\,\,\, 0\leq i \leq k-1,
\end{equation}
\begin{equation}
\tilde{D}_{\text{min}}^{(i)} = D_{\text{min}}^{(i+1)}\,\,\,\, k\leq i \leq \ell-2.
\end{equation}
The operation that created matrix $\tilde{G}$ from $G$ is referred to as shortening $G$ on $(k,j)$.
\end{lemma}
Korada \textit{et al.} \cite{Korada} used Lemma \ref{lem:Shortn} to obtain linear kernels based on $\ell=31$ BCH matrix. Specifically, they were able to derive row $\#2$ in Table \ref{tbl:errorExp}, which was proven to be the maximal  exponent for linear kernels of $\ell\leq 16$ dimensions.

We begin by considering the following matrix $G$, which generates a $[24,5,12]$ sub-code of the extended Golay code $\mathcal{C}_{24}$.
\begin{align*}
G=\left[ \begin{array}{c}
101010101010101010101010\\
110000111100001111000011\\
111100001111000011110000\\
111111110000000011111111\\
111111111111111100000000
\end{array}
\right]
\end{align*}
The partial distance profile of this sub-code is $\left(12,12,12,16,16\right)$. Hence, by extending $G$ to a generator matrix of the extended
Golay code, we will obtain a partial distance profile $\left(8,8,8,8,8,8,8,12,12,12,16,16\right)$. Note that the all-ones word is not spanned by $G$. Therefore, by adding it to $G$ we obtain the generating matrix $G'$, which generates a $[24,6,8]$ sub-code of $\mathcal{C}_{24}$.
\begin{align*}
G'=\left[ \begin{array}{c}
111111111111111111111111\\
101010101010101010101010\\
110000111100001111000011\\
111100001111000011110000\\
111111110000000011111111\\
111111111111111100000000
\end{array}
\right],
\end{align*}
We note, that all the columns of $G'$ are distinct and it contains the all-ones row. Therefore, $G'$ is a parity check matrix of a $[24,18,4]$ code which is a shortening of $[32,26,4]$ extended Hamming code. Because $\mathcal{C}_{24}$ is self-dual, $G'$ is a parity check matrix of a code $C'$ of which $\mathcal{C}_{24}$ is a sub-code. So, the partial distance sequence of the code $C'$ is $\left(4,4,4,4,4,4,8,8,8,8,8,8,8,12,12,12,16,16\right)$.
All the codewords in $C'$ are of even Hamming weight, therefore we can complete the generator matrix of $C'$ to form a $24\times24$ matrix $A$ which has $G$ on its last $5$ rows, and with partial distance sequence $\left(1,2,2,2,2,2,4,4,4,4,4,4,8,8,8,8,8,8,8,12,12,12,16,16\right)$.
The exponent of $B$ is $0.51577$, thus $E_{24}\geq 0.51577$.

The codewords of weight $4$ in $C'$ do not contain all the words of length $24$ and weight $3$. This can be seen, by observing that not all the summations of three columns of $G'$ generate one of the remaining columns of $G'$. For example, the summation of columns $0,8,17$ (from the left) results in $[1     0     1     1     0     0]^T$ which is not one of the columns of $G'$, and this means that the $24$ length word ${\bf v}$ having ones only in indices $0,8,17$ is not contained in any $4$ length codeword of $C'$. Using this, we generate a $25\times 25$ generating matrix $A'$ in the following way. We augment to $A$ a zero column to the right, and insert the vector $[ {\bf v} \,\,\, 1]$ between rows $6$ and $7$ of the augmented matrix. The matrix $A'$ has the following partial distance profile $\left(1,2,2,2,2,2,4,4,4,4,4,4,4,8,8,8,8,8,8,8,12,12,12,16,16\right)$, and therefore its exponent is $0.50608$, so $E_{25}\geq 0.50608$.

By Shortening matrix $A$ on $(24,k)$ where $k\in[16]_{-}$, we obtain by Lemma \ref{lem:Shortn} a $23\times 23$ matrix with partial distance profile $\left(1,2,2,2,2,2,4,4,4,4,4,4,8,8,8,8,8,8,8,12,12,12,16\right)$. Therefore, $E_{23}\geq 0.50705$.

By Shortening matrix $A$ on $(23,0)$ and then shortening the resultant matrix on $(19,0)$, we obtain by Lemma \ref{lem:Shortn} a $22\times 22$ matrix with partial distance profile $\left(1,2,2,2,2,2,4,4,4,4,4,4,8,8,8,8,8,8,8,12,12,16\right)$. Therefore, $E_{22}\geq 0.50118$.
Shortening matrix $A$ on $(23,0)$ and then shortening the resultant matrix on $(3,11)$, we obtain a matrix that has the following sub-matrix for its last three rows.
\begin{align*}
\tilde{A}=\left[ \begin{array}{c}
1010101010110110101010\\
0111100110001100111100\\
0001111111100000001111
\end{array}
\right].
\end{align*}
Since column $15$ of $\tilde{A}$ is a zero column, we can shorten on this column, thereby obtaining a generating matrix with a distance profile $\left(1,2,2,2,2,2,4,4,4,4,4,4,8,8,8,8,8,8,12,12, 12\right)$, so $E_{21}\geq 0.49604$.

Consider matrix $B$, which is a $12\times 18$ generator matrix of an $[18,12,4]$ code with partial distance profile $(4,4,4,4,6,6,8,8,8,8,8,16)$. Clearly, the code that $B$ generates is a sub-code of $[18,17,2]$, so $B$ can be completed to an $18\times18$ matrix with the following partial distance profile $(1,2,2,2,2,2,4,4,4,4,6,6,8,8,8,8,8,16)$ and therefore $E_{18}\geq 0.49521$.
\begin{align*}
B=\left[ \begin{array}{c}
0	0	0	0	0	0	0	0	0	0	1	0	0	0	0	 1	1	1\\
0	1	0	0	1	0	0	0	0	0	1	0	1	1	1	 0	1	1\\
0	0	0	0	0	0	0	0	1	0	0	1	0	1	1	 0	0	0\\
0	0	0	0	0	1	1	0	1	0	0	1	0	0	0	 0	0	0\\
1	0	0	1	1	0	0	1	0	0	0	0	0	0	0	 0	1	1\\
0	1	1	0	0	0	0	0	0	1	0	1	0	0	1	 1	0	0\\
0	1	1	1	1	0	0	0	0	1	0	0	0	1	0	 0	1	1\\
0	1	0	1	0	1	0	1	0	1	0	1	0	1	0	 1	0	0\\
0	0	1	1	0	0	1	1	0	0	1	1	0	0	1	 1	0	0\\
0	0	0	0	1	1	1	1	0	0	0	0	1	1	1	 1	0	0\\
0	0	0	0	0	0	0	0	1	1	1	1	1	1	1	 1	0	0\\
1	1	1	1	1	1	1	1	1	1	1	1	1	1	1	 1	0	0
\end{array}
\right].
\end{align*}

Consider the matrix $F$, which is a $13\times 19$ generator matrix of an $[19,13,4]$ code with partial distance profile $(4,4,4,4,4,6,8,8,8,8,8,8,16)$. This code is a sub-code of $[19,18,2]$ (the single parity check code), so $F$ can be completed to a $19\times 19$ matrix with the following partial distance sequence $(1,2,2,2,2,2,4,4,4,4,4,6,8,8,8,8,8,8,16)$, and therefore $E_{19}\geq 0.49045$.
\begin{align*}
F=\left[ \begin{array}{c}
1	1	0	0	0	0	0	0	0	0	0	0	0	0	0	 0	1	0	 1\\
0	0	0	0	0	0	0	0	0	0	1	0	0	0	0	 1	1	1	 0\\
0	1	0	0	1	0	0	0	0	0	1	0	1	1	1	 0	1	1	 0\\
0	0	0	0	0	0	0	0	1	0	0	1	0	1	1	 0	0	0	 0\\
0	0	0	0	0	1	1	0	1	0	0	1	0	0	0	 0	0	1	 1\\
1	0	0	1	1	0	0	1	0	0	0	0	0	0	0	 0	1	1	 0\\
0	1	1	0	0	0	0	0	0	1	0	1	0	0	1	 1	0	1	 1\\
0	1	1	1	1	0	0	0	0	1	0	0	0	1	0	 0	1	1	 0\\
0	1	0	1	0	1	0	1	0	1	0	1	0	1	0	 1	0	0	 0\\
0	0	1	1	0	0	1	1	0	0	1	1	0	0	1	 1	0	0	 0\\
0	0	0	0	1	1	1	1	0	0	0	0	1	1	1	 1	0	0	 0\\
0	0	0	0	0	0	0	0	1	1	1	1	1	1	1	 1	0	0	 0\\
1	1	1	1	1	1	1	1	1	1	1	1	1	1	1	 1	0	0	 0
\end{array}
\right].
\end{align*}

We note that row $\#1$ of Table \ref{tbl:UBEl} may also be obtained by a method of decomposition and puncturing of the the Golay code $\mathcal{C}_{24}$. This is done by the well-known generation of the Nordstrom-Robinson code from $\mathcal{C}_{24}$.

\section{Summary and Conclusions}

The objective of this study was to construct and analyze  polar code kernels which have better error correcting performance than the standard $(u+v,v)$ polar codes. The performance is manifested by the polar code exponent. Using known code decompositions we were able to construct three new  kernels of dimensions $14,15$ and $16$. These kernels achieve the upper-bound on the exponent per their size which makes them optimal in the sense of the exponent. Furthermore, the kernels are non-linear, which gives the first example  of the advantage that non-linear kernels have upon the linear ones.

The upper-bound we developed turned out to be tight for  $\ell \in [16]\backslash\{12,13\}$, because there exist polar code constructions with exponents that achieve the bound. However, for the other cases there is no evidence if the bound is tight. The linear kernels, that we developed in Section \ref{sec:ExmpofModifiying}, have the largest exponent that we know per their kernel size ($17 \leq \ell \leq 25$), however they do not achieve the upper-bound. We summarize  in Table \ref{tbl:UBSummary} the current knowledge on the optimal exponents of binary kernels for dimensions $5\leq\ell\leq25$. Note that the lower bounds for $\ell\leq 16$ were derived by decompositions from \cite[Table 5]{LitsynTblBinaryCodes}.

It should be emphasized that by using non-binary kernels, it is possible to  get better exponents \cite{MoriandTanka3}. There is an essential loss, when using non-binary code decomposition for designing binary kernels. It seems that if we allow the inputs of the kernel to be from different alphabet sizes, we may gain an additional improvement. This  idea is further explored in a sequel paper \cite{Presman2011}.

\begin{table}
\center
\begin{tabular}{|c|c||c|c|}
   \hline
 ${\ell}$&$E_{\ell}$  & ${\ell}$&$E_{\ell}$\\
  \hline
  5 & $0.43067$ & 6 &  $0.45132$ \\
  7 & $0.45798$& 8 & $0.5$\\
  9 &  $0.46162$ & 10 & $0.46915$\\
  11& $0.47748$ & 12  &   $0.49210-0.49605$ \\
   13 &  $0.49380-0.50049$ & 14 &  $0.50194$ \\
   15 &   $0.50773$& 16 &  $0.52742$ \\
   17 &  $0.49175^{\clubsuit}-0.50447$& 18 &  $0.49521-0.50925$ \\
   19 &  $0.49045-0.51475$&20 &  $0.49659^{\clubsuit}-0.52190$\\
  21&  $0.49604-0.52554$&22&$0.50118-0.52317$ \\
  23&  $0.50705-0.52739$&24 & $0.51577-0.53362$\\
   25 & $0.50608-0.53633$& &\\
  \hline
\end{tabular}
 \normalsize
\caption{The optimal exponent $E_{\ell}$ for binary kernels of different sizes $\ell$. When the exact value of $E_{\ell}$ is not known yet, an interval of values is given. Lower-bounds from \cite[Table I]{Korada} are indicated by $\clubsuit$. }\label{tbl:UBSummary}
\end{table}

\section*{Acknowledgements}
The authors would like to thank the editor and the anonymous reviewers for their helpful and constructive comments that contributed to improving the final version of the paper.

\ifloguseIEEEConf
    \appendix
\else
    \section*{Appendix}
\fi
In this appendix, we give  details on the decompositions enumerated in Table \ref{tbl:errorExp}. All of the decompositions are coset decompositions, so we only need to specify the sub-code representatives.

\subsubsection*{\#1)$(16,16,1)-(16,15,2)-(16,11,4)-(16,8,6)-(16,5,8)-(16,1,16)$}
The sub-code representatives are  $(16,15,2)$  single parity check code,  $(16,11,4)$ extended Hamming code, $(16,8,6)$ Nordstrom-Robinson code, $(16,5,8)$ first order Reed-Muller code, $(16,1,16)$ repetition code.
\subsubsection*{\#2)$(16,16,1)-(16,15,2)-(16,11,4)-(16,7,6)-(16,5,8)-(16,1,16)$}
The sub-code representatives are  $(16,15,2)$ - single parity check
code,  $(16,11,4)$ - extended Hamming code, $(16,7,6)$ -   extended
$2$-error correcting BCH code, $(16,5,8)$-   first-order Reed-Muller
code, $(16,1,16)$ - repetition code.
\subsubsection*{\#3)$(15,15,1)-(15,14,2)-(15,10,4)-(15,7,6)-(15,4,8)$}
The sub-code representatives are $(15,14,2)$ - single parity check
code,  $(15,10,4)$ - shortened extended Hamming code, $(15,7,6)$ - shortened
Nordstrom-Robinson code, $(15,4,8)$ - shortened first order
Reed-Muller code.
\subsubsection*{\#4)$(14,14,1)-(14,13,2)-(14,9,4)-(14,6,6)-(14,3,8)$}
The sub-code representatives are $(14,13,2)$ - single parity check
code,  $(14,9,4)$ - twice shortened extended Hamming code,
$(14,6,6)$ - twice shortened Nordstrom-Robinson code, $(14,3,8)$ -
twice shortened first order Reed-Muller code.

\subsubsection*{Explicit Encoding of Decomposition $\#1$}
For decomposition $\#1$ we elaborate on the kernel mapping function $g(\cdot):\left\{0,1\right\}^{16}\rightarrow\left\{0,1\right\}^{16}$. In order to do this, we use Table \ref{tbl:elabCodeDecomp}.
The third column from the left determines whether the vectors on the second column are all the coset vectors (if they do not form a linear space) or just the basis for the space of coset vectors (if they form a linear space). The fourth and the fifth columns determine the stage of the code decomposition these vectors belong to;  the "main code" is decomposed to cosets of the "sub-code" (each coset is generated by adding a different coset vector from the set specified by column 2 to the sub-code). The entry corresponding to indices $8-10$ is taken from  \cite{LitsynFastDecoding}.

We now describe the encoding process. Let ${\bf u}_{0}^{15}$ be a binary vector. The indices of the vector are partitioned into subsets according to the first column of the table.  For each subset, the corresponding sub-vector of $\bf u$ is mapped to a coset vector. The mapping can be arbitrary, however when the coset vectors form a linear space, we usually prefer to multiply the corresponding sub-vector by a generating matrix, where the rows are the vectors in the "coset vectors" column. To get the value of $g({\bf u})$, we add-up the six coset vectors we got from the last step. Note that using this mapping definition, it is also easy to derive the mapping functions corresponding to decompositions $\#3$ and $\#4$ as well.

Hammons \textit{et al.} showed that some famous binary non-linear codes can be represented as binary images under the Gray mapping of linear codes over the $\mathbb{Z}_4$ ring \cite{Hammons1994}. In particular, The Nordstrom-Robinson code was proven to hold this property. Following this approach, Table \ref{tbl:elabCodeDecomp} decomposition can be represented as linear decomposition over  $\mathbb{Z}_4$ using the generating matrix $G_{\#1}$.
\begin{equation}\label{eq:Z4decomposition16}
G_{\#1}=\left[ \begin{array}{c}
\textbf{0	0	0	0	0	0	0	1}\\
\hdashline
\textbf{0	0	0	1	0	0	0	1}\\
\textbf{0	0	0	0	0	1	0	1}\\
\textbf{0	0	0	0	0	0	1	1}\\
\textbf{0	0	0	0	0	0	0	2}\\
\hdashline
\textbf{0	1	0	1	0	1	0	1}\\
\textbf{0	0	1	1	0	0	1	1}\\
\textbf{0	0	0	0	1	1	1	1}\\
\hdashline
\textbf{0	1	2	1	0	3	0	1}\\
\textbf{0	0	1	3	0	2	1	1}\\
\textbf{0	0	2	0	1	1	3	1}\\
\hdashline
\textbf{1	1	1	1	1	1	1	1}\\
\textbf{0	2	0	2	0	2	0	2}\\
\textbf{0	0	2	2	0	0	2	2}\\
\textbf{0	0	0	0	2	2	2	2}\\
\hdashline
\textbf{2	2	2	2	2	2	2	2}\\
\end{array}
\right].
\end{equation}
The entries of $G_{\#1}$ are from $\mathbb{Z}_4$ (indicated here by bold typeface), and the Gray mapping is defined as $\textbf{0}\rightarrow 00$; $\textbf{1}\rightarrow 01$; $\textbf{2}\rightarrow 11$; $\textbf{3}\rightarrow 10$.
Generating the codewords ${\bf x} \in \mathbb{Z}_4^8$ is performed by  multiplying a $16$ length binary vector  ${\bf u}$ (referred to as the input vector) by $G_{\#1}$, i.e.  ${\bf x} = {\bf u}\cdot  G_{\#1}$ where ${\bf u}\in \{0,1\}^{16}$. The dashed-lines in (\ref{eq:Z4decomposition16}) correspond to the different steps of the decomposition in Table \ref{tbl:elabCodeDecomp}.

Having a binary information vector used in a $\mathbb{Z}_4$ linear code definition is rather untraditional, and was employed here to support the binary decomposition representation. We now give some hints on how this generating matrix can be transformed into a $\mathbb{Z}_4$ canonical matrix form defined in \cite[Section II.A]{Hammons1994}.  Denote the rows of $G_{\#1}$ by ${\boldsymbol \gamma}_i$ where $i\in [16]_{-}$.  It can be easily seen that ${\boldsymbol \gamma}_i = 2\cdot{\boldsymbol \gamma}_{i+4}$  for $8\leq i\leq 11$. Furthermore rows $\left[{\boldsymbol \gamma}_i\right]_{i=11}^{14}$  form a generating matrix for the first-order Reed-Muller code of which the input vector ${\bf v}_{0}^3$ is in $\mathbb{Z}_4\times \{0,1\}^3$, i.e. the code is of type $4^12^3$ (see  \cite[Section II.A]{Hammons1994} for the definition of code type). Rows  $\left[{\boldsymbol \gamma}_i\right]_{i=8}^{11}$ form a generating matrix for the Nordstrom-Robinson code, of which the input vector ${\bf w}_{0}^3$ is in $\mathbb{Z}_4^4$, i.e. the code is of type $4^4$. By performing rows replacement ${\boldsymbol \gamma}_i  \leftarrowtail \left({\boldsymbol \gamma}_i- {\boldsymbol \gamma}_{i+4}\right) $ where $5 \leq i \leq 7$, it can be easily shown that rows $\left[{\boldsymbol \gamma}_i\right]_{i=5}^{11}$ form a generating matrix for the extended Hamming code which is of type $4^42^3$. The code spanned by the sub-matrix containing rows  $\left[{\boldsymbol \gamma}_i\right]_{i=1}^{11}$ is the single parity check code defined by the following generating matrix over  $\mathbb{Z}_4$

\begin{equation}\label{eq:singleParity}
G_{sp}=\left[ \begin{array}{l   : l}
{\bf I}_7	& \overrightarrow{{\bf 1}}^T	\\[5pt]
\hdashline
\overrightarrow{{\bf 0}} & {\bf 2}\\
\end{array}
\right],
\end{equation}
where ${\bf I}_7$ is the $7\times 7$ identity matrix and $\overrightarrow{\bf 0}$ and $\overrightarrow{\bf 1}$ are, respectively, the all-zeros and the all-ones row vectors of length $7$. As a consequence the single parity check code is a $4^72^1$ type code.

\subsubsection*{SC Decoding}

In this section we briefly cover SC decoding for polar codes. We begin our discussion by considering a polar code generated by a single application of an $\ell$-dimensions kernel. Let ${\bf u}_{0}^{\ell-1}$ and ${\bf x}_{0}^{\ell-1}$, be two binary vectors such that $g({\bf u})={\bf x}$, where $g(\cdot)$ is an $\ell$-dimensions kernel that was defined in Section \ref{sec:CodesDecom}. The codeword
${\bf x}$ is transmitted over $\ell$ copies of the binary memoryless channel $\mathcal{W}$ and  the channel output vector $\bf y$ is received. As was noted in  Section \ref{sec:CodesDecom}, an SC decoder sequentially decides on the most likely sub-code to which the transmitted codeword ${\bf x}$ belongs to, given its noisy observation $\bf y$. The sub-codes of the decomposition are identified by the information vector ${\bf u}$. On step number $i\in \left[\ell\right]_{-}$ of the SC decoding algorithm, we decide on the non-frozen bit $u_i$ given the decisions on the previous symbols (${\hat{\bf u}}_{0}^{i-1}$). In order to do so we have to calculate two likelihoods:
\begin{equation}\label{eq:scDecoding}
    W_{\ell}^{(i)} \left({\bf y}, {\hat{\bf u}}_{0}^{i-1} \left| u_{i}=b \right.  \right) \triangleq \frac{1}{2^{\ell-1}}\sum_{{\bf u}_{i+1}^{\ell-1}\in \{0,1\}^{\ell-i-1}}W_{\ell}\left( {\bf y}\left| {\hat{\bf u}}_{0}^{i-1},b, {{\bf u}}_{i+1}^{\ell-1}\right.\right) =
\end{equation}
$$
= \frac{1}{2^{\ell-1}}\sum_{{\bf x}\in T^{\left(\hat{\bf u}_{0}^{i-1}\cdot b \right)}_{i+1}}\prod_{i=0}^{\ell-1}W\left( y_i\left| x_i\right.\right),\,\,\,\,\,\,  b \in \{0,1\},
$$
where $W_{\ell}\left(\cdot|\cdot\right)$ is the transition function of the channel ${\bf u}_{0}^{\ell-1}\rightarrow {\bf y}_{0}^{\ell-1}$ and $W\left(\cdot|\cdot\right)$ is the transition function of the channels $x_i\rightarrow y_i$, $i\in \left[\ell\right]_{-}$.
We then can decide on the value of $u_i$ by employing the maximum likelihood (ML) rule:
$$
\hat{u}_i=\left\{
  \begin{array}{ll}
    0, & \hbox{$ W_{\ell}^{(i)} \left({\bf y}, {\hat{\bf u}}_{0}^{i-1} \left| u_{i}=0 \right.  \right) > W_{\ell}^{(i)} \left({\bf y}, {\hat{\bf u}}_{0}^{i-1} \left| u_{i}=1 \right.  \right) $;} \\
    1, & \hbox{otherwise.}
  \end{array}
\right.
$$
After applying the ML decision rule on $u_i$, the SC decoder proceeds to the next step.

The straight-forward calculation of the likelihoods performed on decoding step
$\#i$ ($i\in \left[\ell\right]_{-}$) based on (\ref{eq:scDecoding}) requires $2\cdot\left(2^{\ell-i-1}-1\right)$ additions and $2^{\ell-i}\cdot(\ell-1)$ multiplications. For linear kernels it is possible to perform trellis decoding based on the zero-coset's parity check matrix. In this way the number of additions is $\leq \ell \cdot 2^{i+1}$ and the number of multiplications is $\leq \ell \cdot 2^{i+2}$. These bounds do not take into account the fact that some paths in the trellis may be skipped and that some of the nodes in the trellis have input degree $<2$. Note further that due to numerical stability it is preferable to use log-likelihoods instead of likelihoods in the decoding algorithm implementation. In this case the number of likelihoods multiplications should be regarded as the number of log-likelihoods additions. The number of additions should be understood as the number of $\max^{\star}(\cdot,\cdot)$ operations, where $\max^{\star}\left(\alpha_0,\alpha_1\right)\triangleq\max\{\alpha_0,\alpha_1\} + \log\left(1+\exp\left\{|\alpha_1-\alpha_0|\right\}\right)$.

SC decoding of a length $N=\ell^n$ bits code, involves recursive applications of the likelihood calculation for a single kernel (see e.g. \cite{Presman2012}). In order to calculate the total number of operations employed in the SC algorithm for a code of length $N$ bits, we need to take into account the number of occurrences of each kernel decoding step in the algorithm. An upper-bound on this quantity can be easily derived by counting the number
of kernels in the code structure of each polar code, which is $\frac{N}{\ell}\cdot \log_{\ell}N$. Consequently the time complexity of for SC decoding  of a general code is $O\left(2^{\ell}\cdot N\cdot\ \log_{\ell} N \right)$. For linear codes the time complexity may be reduced to  $O\left(2^{\ell/2}\cdot N\cdot\ \log_{\ell} N \right)$ by incorporating trellis decoding.

For our proposed non-linear kernels, trellis can still be used to reduce the decoding complexity because they can be represented as   linear decompositions over  $\mathbb{Z}_4$. Finding the parity check matrix corresponding to the different sub-codes can be done using equation (2) in  \cite{Hammons1994}.

\begin{table}
\scriptsize
\center
\begin{tabular}{| c|c|c|c|c|}
  \hline
  input   & coset vectors &  coset vectors  & main code & sub-code
  \\
             vector  &              &    form a &      &
               \\
                 indices     &              &         linear space?  &   &
               \\
  \hline
  $0$ & $[0	0	0	0	0	0	0	0	0	0	0	0	0	 0	0	 1
]$ & yes & $(16,16,1)$ & $(16,15,2)$ \\
\hline
  $1-4$ & $[0	0	0	0	0	0	0	1	0	0	0	0	 0	0	0	 1
]$ & yes & $(16,15,2)$ & $(16,11,4)$ \\
    & $[0	0	0	0	0	0	0	0	0	0	0	1	0	 0	0	 1
]$ &   &   &   \\
    & $[0	0	0	0	0	0	0	0	0	0	0	0	0	 1	0	 1
]$ &   &   &   \\
    &  $[0	0	0	0	0	0	0	0	0	0	0	0	0	 0	1	 1
]$ &   &   &   \\
\hline
  $5-7$ & $[0	0	0	1	0	0	0	1	0	0	0	1	 0	0	0	 1
]$ & yes & $(16,11,4)$ & $(16,8,6)$ \\
   & $[0	0	0	0	0	1	0	1	0	0	0	0	0	 1	0	 1]$
 &   &   &   \\
   & $[0	0	0	0	0	0	0	0	0	1	0	1	0	 1	0	 1
]$ &   &   &   \\
\hline
  $8-10$ & $[0	0	0	0	0	0	0	0	0	0	0	0	 0	0	0	 0
] $& no &  $(16,8,6)$ & $(16,5,8)$ \\
  & $[0	0	0	0	0	0	1	1	0	1	0	1	0	1	 1	0
]$ &   &   &   \\

    & $[0	0	0	1	0	0	0	1	0	1	0	0	1	 0	1	 1
]$ &   &   &   \\
       & $[0	0	0	1	0	0	1	0	0	0	1	0	 1	1	1	 0
]$ &   &   &   \\
   & $[0	0	0	1	0	1	1	1	0	0	0	1	1	 0	0	 0
]$ &   &   &   \\
  & $[0	0	0	0	0	1	1	0	0	0	1	1	0	1	 0	1
]$ &   &   &   \\
    & $[0	0	0	1	0	1	0	0	0	1	1	1	0	 0	1	 0
]$ &   &   &   \\
  & $[0	0	0	0	0	1	0	1	0	1	1	0	1	1	 0	0
]$ &   &   &   \\
\hline
  $11-14$ & $[0	1	0	1	0	1	0	1	0	1	0	1	 0	1	0	 1
]$ & yes & $(16,5,8)$ & $(16,1,16)$ \\
    & $[0	0	1	1	0	0	1	1	0	0	1	1	0	 0	1	 1
]$ &  &  &  \\
   & $[0	0	0	0	1	1	1	1	0	0	0	0	1	 1	1	 1
]$ &   &   &   \\
    & $[0	0	0	0	0	0	0	0	1	1	1	1	1	 1	1	 1
]$ &   &   &   \\
\hline
  $15$ & $[1	1	1	1	1	1	1	1	1	1	1	1	 1	1	1	 1]$ &  yes &  $(16,1,16)$ & -  \\
  \hline
\end{tabular}
\normalsize
\caption{Coset vectors for code decomposition $\#1$.   }\label{tbl:elabCodeDecomp}
\end{table}

\bibliographystyle{IEEEtran}
\bibliography{IEEEabrv,bibTexPolar}

\end{document}